\documentclass[a4paper, 12pt, fleqn]{article}
\usepackage[hmargin={1in, 1in}, vmargin={1in, 1in}]{geometry}
\usepackage[numbers, sort&compress]{natbib} 
\usepackage{color}        
\usepackage{graphicx}
\usepackage{amssymb}
\usepackage{epstopdf}
\usepackage{caption}	
\begin{document}


\title{\begin{flushleft} Measurement and physical interpretation of the mean motion of turbulent density patterns detected by the BES system on MAST\end{flushleft}}
\author{Y-c Ghim$^{1,2}$, A R Field$^2$, D Dunai$^3$, S Zoletnik$^3$, L Bard\'{o}czi$^3$, A A Schekochihin$^1$, \\ and the MAST Team$^2$}
\date{\begin{flushleft}
\textit{$^1$Rudolf Peierls Centre for Theoretical Physics, University of Oxford, Oxford, OX1 3NP, United Kingdom\newline
           $^2$EURATOM/CCFE Fusion Association, Culham Science Centre, Abingdon, OX14 3DB, United Kingdom\newline
           $^3$Wigner Research Centre for Physics, Association EURATOM/HAS, P.O. Box 49, H-1525, Budapest, Hungary}\newline
           E-mail: \texttt{y.kim1@physics.ox.ac.uk}\newline
        \end{flushleft}}                                          

\maketitle
\thispagestyle{headings} 

\newpage
\markright{Motion of turbulent density patterns detected by the BES system on MAST}

\noindent
{\bf Abstract.}  
The mean motion of turbulent patterns detected by a two-dimensional (2D) beam emission spectroscopy (BES) diagnostic on the Mega Amp Spherical Tokamak (MAST) is determined using a cross-correlation time delay (CCTD) method.  Statistical reliability of the method is studied by means of synthetic data analysis.  The experimental measurements on MAST indicate that the apparent mean poloidal motion of the turbulent density patterns in the lab frame arises because the longest correlation direction of the patterns (parallel to the local background magnetic fields) is not parallel to the direction of the fastest mean plasma flows (usually toroidal when strong neutral beam injection is present).   The experimental measurements are consistent with the mean motion of plasma being toroidal.  The sum of all other contributions (mean poloidal plasma flow, phase velocity of the density patterns in the plasma frame, non-linear effects, etc.) to the apparent mean poloidal velocity of the density patterns is found to be negligible.  These results hold in all investigated L-mode, H-mode and internal transport barrier (ITB) discharges.  The one exception is a high-poloidal-beta (the ratio of the plasma pressure to the poloidal magnetic field energy density) discharge, where a large magnetic island exists.  In this case BES detects very little motion.  This effect is currently theoretically unexplained.  
\newline\newline
(Some figures in this article are in colour only in the electronic version)
\newpage


\section{Introduction}\label{sec:intro}
\noindent
It is now widely accepted that turbulent transport in magnetically confined fusion plasmas can exceed the irreducible level of neoclassical transport by an order of magnitude or more \cite{carreras_ieee_1997}. However, both theoretical and experimental works of the past two decades \cite{highcock_pop_2011, highcock_prl_2010, waltz_pop_1994, waltz_pop_1997, dimits_nf_2001, kinsey_pop_2005, camenen_pop_2009, roach_ppcf_2009, barnes_prl_2011, burrell_pop_1997, connor_2004, mantica_prl_2009, mantical_prl_2011, devries_nf_2009} suggest that sheared $\vec{E}\times\vec{B}$ flows can moderate such anomalous transport and hence improve the performance of magnetically confined fusion plasmas. 
\newline\indent 
With the aim of characterizing the microscale plasma turbulence and searching for correlations between it and the background plasma characteristics, a two-dimensional (8 radial $\times$ 4 poloidal channels) beam emission spectroscopy (2D BES) system \cite{field_rsi_2011} has been installed on the Mega Amp Spherical Tokamak (MAST).  It is able to measure density fluctuations at scales above the ion Larmor radius $\rho_i$, viz., $k_\perp\rho_i<1$, where $k_\perp$ is the wavenumber perpendicular to the magnetic field.  The 2D BES view plane lies on a radial-poloidal plane at a fixed toroidal location.  Following the detected turbulent density patterns on this view plane allows one to determine their mean velocity in the radial and poloidal directions.  Typically, there are no significant mean plasma flows in the radial direction in a tokamak, whereas considerable apparent poloidal motion is detected by the 2D BES system.
\newline\indent
In this paper, we show that this apparent poloidal motion is primarily due to fact that the direction of the longest correlation of the turbulent density patterns is not parallel to that of the dominant mean plasma flows.  The BES measurements are shown to be consistent with a dominantly toroidal mean flow; the poloidal flows are of the order of the diamagnetic velocities.  These results are obtained using the cross-correlation time delay (CCTD) method, which is a frequently used statistical technique to determine the apparent velocity of density patterns \cite{durst_rsi_1992, cosby_master_1992}.  We also investigate the method itself thoroughly to determine the statistical uncertainties of the technique. This is done by generating synthetic 2D BES data with random Gaussian density patterns calculated on a graphical processing unit (GPU) card using CUDA (Compute Unified Device Architecture) programming.
\newline\indent
The paper is organized as follows.  In section \ref{sec:quantity 2D BES measures}, we explain what is measured directly by the 2D BES system, and how the apparent velocity of turbulent density patterns can be inferred from this data.  We also show what physical effects contribute to the apparent velocity calculated by the CCTD method.  In section \ref{sec:exp_results}, we present the experimental results with the aim of identifying the main cause of apparent motion of density patterns measured by the 2D BES system.  Our conclusions are presented in section \ref{sec:conc}.  For the readers who are interested in the statistical technique employed in this paper to determine the velocity of the density patterns, the cross-correlation time delay (CCTD) method and its statistical reliability are studied in \ref{sec:the_cctd_method} using synthetically generated 2D BES data (described in \ref{sec:gen_syn_data}).

\section{What is measured by the 2D BES system}\label{sec:quantity 2D BES measures}
\noindent
The 2D BES system on MAST utilizes an avalanche photodiode (APD) array camera \cite{dunai_rsi_2010} with 8 radial and 4 poloidal channels, which have an active area of $1.6 \times 1.6\:mm^2$ each. It measures the Doppler-shifted $D_{\alpha}$ emission from the collisionally excited neutral beam atoms (deuterium) with a temporal resolution of 0.5 $\mu sec$. The optical system is designed so that the 2D BES system can scan radially along the neutral beam, whose $1/e$ half-width is 8 $cm$, while the optical focal point follows the axis of the beam.  The nominal location of the BES system, i.e., where the optical line-of-sight (LoS) is best aligned with the local magnetic field, is at major radius $R=1.2\:m$.  At this location, a magnification factor of 8.7 at the axis of the beam results in each channel observing an area of $1.5 \times 1.5\:cm^2$ with $2\:cm$ separation between the centres of adjacent channels.   The angle between the LoS of the 2D BES system and the neutral beam with the injection energy of $60-70 keV$ results in a Doppler shift of the $D_{\alpha}$ emission approximately $3\:nm$ to the red from the background $D_{\alpha}$.  The background $D_{\alpha}$ can be removed with a suitable optical filter, and so only the detected $D_{\alpha}$ emission by the 2D BES system comes from the neutral beam, hence the locality for measurement to the beam.  Aligning the LoS parallel to the local magnetic field at the intersection of the LoS and the neutral heating beam helps minimize the degradation of the spatial resolution.  A more detailed description of the 2D BES system on MAST with possible sources of some losses of spatial locality can be found elsewhere \cite{field_rsi_2011, field_rsi_2009}.

\subsection{Plasma density fluctuations}\label{sec:measure dens fluc}
\noindent
The measured intensity of the $D_{\alpha}$ beam emission is directly related to the background plasma density because the latter is the cause of the excitation of the neutral beam atoms. The beam atoms can be excited by electrons, ions and impurities, but for the injection energy greater than $40\:keV$, the electron contribution can be ignored \cite{fonck_rsi_1990}.  The fluctuating part of the plasma (ion) density $\delta n$ can be determined according to
\begin{equation}\label{eq:photon_dens_relation}
\frac{\delta n}{n} = \frac{1}{\beta}\frac{\delta I}{I},
\end{equation}
where $n$ is the mean plasma density, $\delta I$ and $I$ denote the fluctuating and mean parts of the photon intensity, respectively, and $\beta$ is calculated based on the ADAS (Atomic Data and Analysis Structure) database \cite{summers_adas_2004}.  $\beta$ is a weak function of the background plasma density ranging approximately from $1/3$ to $1/2$.
\newline\indent
Thus, the 2D BES system on MAST directly measures fluctuations of plasma density in the poloidal-radial plane at a fixed toroidal location.  The spatial resolution of the system is reduced by smearing due to the effects of field-line curvature, observation geometry, finite lifetime of the excited neutral-beam atoms, and the attenuation and divergence of the beam.  These effects must all be taken into account in the calculation of the point spread functions (PSFs) of the detectors comprising the 2D BES system.  A detailed calculation shows that $\sim2-3\:cm$ radial resolution and $\sim1-5\:cm$ poloidal resolution are achievable, depending somewhat on the radial viewing locations \cite{ghim_rsi_2010}.

\subsection{Velocity of density patterns}\label{sec:vel measure}
From the time-dependent 2D measurement of density fluctuations, one can infer the apparent velocity of the density patterns.  This has been the subject of much attention \cite{durst_rsi_1992, shafer_pop_2012, zweben_ppcf_2012, tal_pop_2011, xu_ppcf_2011, zweben_pop_2006, jakubowski_prl_2002} in the hope that this velocity can be related in a more or less straightforward way to the actual plasma flows.  We will first explore how the mean pattern velocity can be determined and then discuss the interpretation of this quantity.

\subsubsection{The CCTD method}\label{subsec:using_the_cctd_method}
\begin{figure}[!t]
\centering
\includegraphics[width=6.5in]{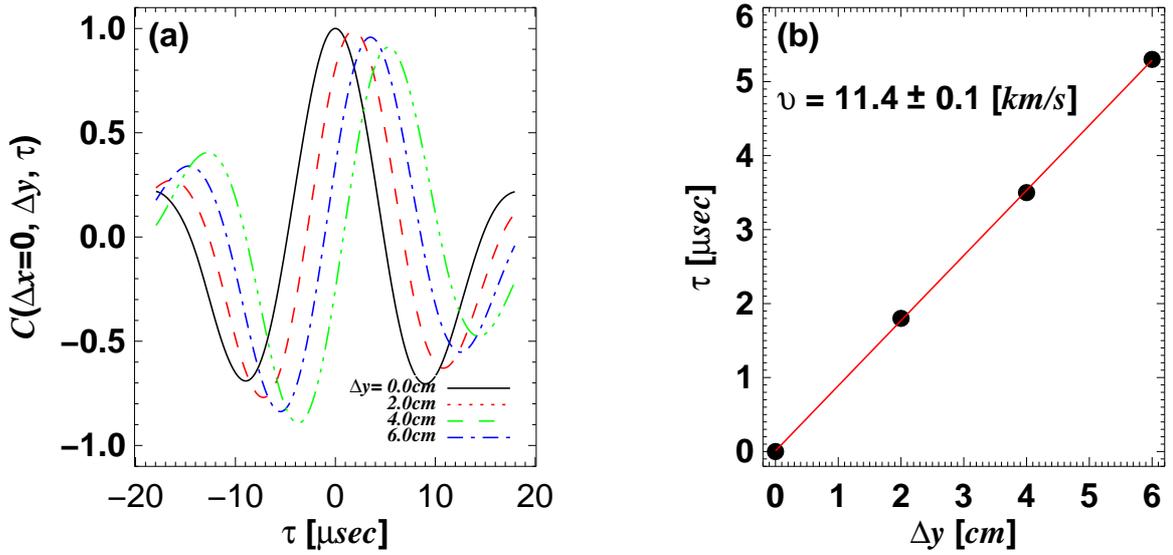}
\caption{\label{fig:cctd_method}(a) Cross-correlation functions calculated using equation (\ref{eq:cc_definition}) for $\Delta y = 0.0\:cm$ (black solid line), $2.0\:cm$ (red dash line), $4.0\:cm$ (blue dash dot line) and $6.0\:cm$ (green dash dot dot line). $\tau_{peak}^{cc}\left(\Delta y\right)$ is the position of maximum of the cross-correlation function. (b) Position of maximum $\tau_{peak}^{cc}\left(\Delta y\right)$ and a linear fit.  The measured velocity is $11.4\pm0.1\:km/s$.}
\end{figure}

\noindent
The CCTD (cross-correlation time delay) method has been widely used to determine the apparent velocities of turbulent density patterns detected by BES systems, and it is well described in \cite{durst_rsi_1992} and \cite{cosby_master_1992}. Here, a brief summary of the method is provided. The normalized fluctuating intensity of the photons, $\hat I\equiv\delta I/I$, measured by a 2D BES system is a function of the radial $x$, vertical (poloidal) $y$ and time $t$ coordinates: $\hat{I}=\hat{I}\left(x, y, t\right)$.  The cross-correlation function of this fluctuating signal is defined as
\begin{equation}\label{eq:cc_definition}
\mathcal{C}\left( \Delta x, \Delta y, \tau \right) = 
\frac{\left\langle \hat{I} \left( x, y, t \right) 
\hat{I} \left( x+\Delta x, y+\Delta y, t+\tau \right)\right\rangle}{\sqrt{\left\langle \hat{I}^2 \left( x, y, t \right) \right\rangle 
\left\langle \hat{I}^2 \left(x+\Delta x, y+\Delta y, t+\tau \right) \right\rangle}},\end{equation}
where $\Delta x$ and $\Delta y$ are the radial and vertical (poloidal) channel separation distances, respectively, $\tau$ is the time lag, and $\left\langle \cdot \right\rangle$ denotes time average defined in \ref{sec:the_cctd_method}.  The apparent poloidal velocity $v_{y}^{BES}$ of the density patterns detected by the 2D BES system can be determined from the time lag $\tau_{peak}^{cc}$ at which the cross-correlation function reaches its maximum for a given $\Delta y$ and $\Delta x=0$\footnote{We concentrate on the apparent mean `poloidal' motion of the density patterns.  Thus, the information about the radial correlations of the 2D BES data is not used in this paper.}.  If a straight line is fitted to the experimentally measured $\tau_{peak}^{cc}\left(\Delta y\right)$, the inverse of its slope is the velocity $v_y^{BES}$.  Although any two poloidally separated channels are sufficient to determine $v_y^{BES}$, using just two channels is insufficient to estimate the uncertainties in the linear fit.  Thus, in this paper, all four available poloidal channels are used to determine these quantities. This assumes that the mean velocity does not change over the time the density patterns take to move past the four poloidal channels and that the lifetime of these patterns is sufficiently long, so the same patterns are observed by all four channels.
\newline\indent
Figure \ref{fig:cctd_method} shows an example of this procedure.  
This example is based on a synthetic data set consisting of Gaussian-shaped random ``eddies'' moving with the poloidal velocity of $10.0\:km/s$, which are then used to produce artificial 2D BES data (see \ref{sec:gen_syn_data} for the description of the synthetic data).  With the four available poloidal channels, cross-correlation functions are calculated using equation (\ref{eq:cc_definition}) and shown in Figure \ref{fig:cctd_method}(a); $\tau_{peak}^{cc}$ is plotted as a function of $\Delta y$ in Figure \ref{fig:cctd_method}(b). The inverse of the slope of a fitted straight line is the velocity $v_y^{BES}$.   Note the slight discrepancy between the actual and CCTD-determined velocities.  The origin and size of this discrepancy are discussed in \ref{sec:the_cctd_method}.

\subsubsection{Physical meaning of the velocity determined by the CCTD method}\label{sec:physcial_meaning_cctd}
\begin{figure}[!t]
\centering
\includegraphics[width=5.0in]{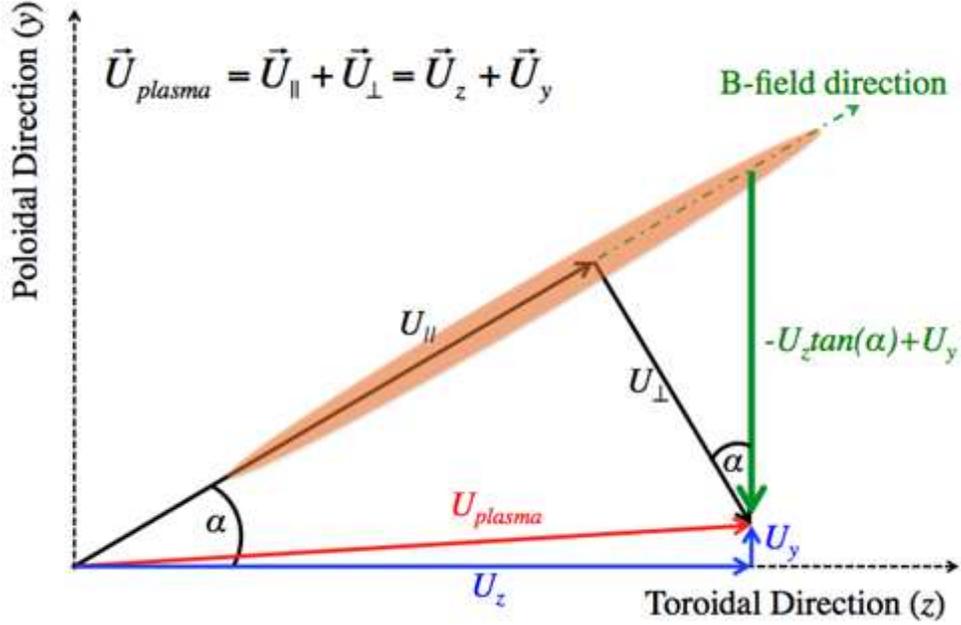}
\caption{\label{fig:barber_pole} Cartoon illustrating how the mean toroidal plasma flow ($U_z)$ induces an apparent mean poloidal motion.  An elongated density pattern (shaded oval) along the magnetic field line (green dash dot) is advected by the toroidal flow (blue arrow).  Because the longest correlation direction of the density pattern is not in the toroidal direction, the apparent mean poloidal flow (green arrow) arises.  The apparent velocity is $-U_z\tan\alpha+U_y\approx -U_z\tan\alpha$, where $\alpha$ is the local magnetic pitch angle.}
\end{figure}

\noindent
Using the described CCTD method, the 2D BES system on MAST is expected to be able to determine $v_y^{BES}$ as has been done previously on TFTR \cite{durst_rsi_1992} and DIII-D \cite{jakubowski_prl_2002} using their BES systems \cite{paul_rsi_1990, mckee_rsi_1999}. However, as McKee {\it et al.} \cite{mckee_pop_2003, mckee_rsi_2003} pointed out, one must distinguish between the poloidal velocity measured by 2D BES system ($v_y^{BES}$) and the actual velocity of the poloidal plasma flow ($U_y$).  
\newline\indent
The mean plasma flow can be decomposed into toroidal ($U_z$) and poloidal ($U_y$) components.  For typical tokamak plasmas where strong neutral beams are injected, $|U_z| \gg |U_y|$ is satisfied as any mean poloidal flows are strongly damped \cite{hirshman_nf_1981, helander_cam_2002}, leaving $U_y$ of the order of the diamagnetic velocity $\sim\rho_*v_{th}$, where $\rho_*=\rho_i/a$, $a$ is the tokamak minor radius, and $v_{th}$ is the ion thermal velocity.  Note that $U_z$ can be on the order of $v_{th}$ for the neutral-beam-heated plasmas.  Thus, $U_y$ can be ignored compared to $U_z$, except possibly in regions with strong pressure gradients.
\newline\indent
As the 2D BES system on MAST observes the density patterns advected by $U_z$, there will be an apparent motion of the patterns in the poloidal direction, as shown in Figure \ref{fig:barber_pole}.
This effect is analogous to the apparent up-down motion of helical strips of a `rotating barber-pole' (cf. \cite{munsat_rsi_2006}).  The magnitude of this apparent velocity can be readily calculated via elementary geometry: namely, we expect the BES system to ``see'', to lowest order in $\rho_*$, 
\begin{equation}\label{eq:simple_v_bes_relation}
v_y^{BES}\approx-U_z\tan\alpha,
\end{equation}
 where $\alpha$ is the pitch angle of the local magnetic field line.
\newline\indent
Equation (\ref{eq:simple_v_bes_relation}) is experimentally verifiable because all three physical quantities are readily obtained by separate diagnostics: $v_y^{BES}$ from the 2D BES system, $U_z$ from the charge exchange recombination spectroscopy (CXRS) system \cite{conway_rsi_2006}, and $\alpha$ either from \texttt{EFIT} equilibrium reconstruction \cite{lao_nf_1985} or directly from the Motional Stark Effect (MSE) system \cite{kuldkepp_rsi_2006, debock_rsi_2008} on MAST.  Although the CXRS system measures the toroidal flow of the $C^{6+}$ ions, the difference between the velocity of the $C^{6+}$ ions and the bulk plasma ions, $D^+$, is predicted to be on the order of $\rho_*$ in a strongly beam-heated plasma \cite{kim_pfb_1991}.  In section \ref{sec:exp_results}, equation (\ref{eq:simple_v_bes_relation}) will be experimentally verified for various types of discharges.  Agreement will indicate consistency of the experiment with the assumptions behind equation (\ref{eq:simple_v_bes_relation}).  Such agreement will indeed be obtained, except in one intriguing case.
\newline\indent
Let us now consider what are the assumptions necessary for equation (\ref{eq:simple_v_bes_relation}) to hold by analysing how the estimated $v_y^{BES}$ depends on actual physical quantities associated with plasma flows and fluctuations in a tokamak.  The cross-correlation function (\ref{eq:cc_definition}) of the normalized fluctuating photon intensity $\hat I$ can, in view of equation (\ref{eq:photon_dens_relation}), be considered proportional to the cross-correlation function of the relative ion density fluctuation $\delta n/n$ (by definition, $\langle\delta n\rangle=0$).  Therefore, the CCTD-determined velocity of the density patterns can be related to the actual physical quantities in a tokamak by invoking the ion continuity equation.  Splitting also the ion velocity into mean and fluctuating parts, $\vec{u}=\vec{U}+\delta\vec{u}$, $\langle\delta\vec{u}\rangle=0$, we have
\begin{equation}\label{eq:continuity_equation_full}
\frac{\partial n}{\partial t}+\frac{\partial\delta n}{\partial t}+\nabla\cdot\left(n\vec{U}+n\delta\vec{u}+\delta n\vec{U}+\delta n\delta\vec{u}\right)=0.
\end{equation}
Averaging this equation and subtracting the averaged equation from (\ref{eq:continuity_equation_full}), we obtain
\begin{equation}\label{eq:continuity_equation}
\frac{\partial\delta n}{\partial t}=-\nabla\cdot\left(n\delta\vec{u}+\delta n\vec{U}+\delta n\delta\vec{u}-\left\langle\delta n\delta\vec{u}\right\rangle\right).
\end{equation}
We will now order various terms in this equation in terms of the small parameter $\rho_*=\rho_i/a$.  
\newline\indent
Assuming that the spatial scale of all mean quantities is $\sim\mathcal{O}(a)$ while the spatial scale of all fluctuating quantities is $\sim\mathcal{O}(\rho_*a)$, and also $\delta n/n \sim \delta u/v_{th} \sim \rho_*$, we get
\begin{eqnarray}\label{eq:continuity_equation_ordered}
\frac{\partial}{\partial t}\frac{\delta n}{n} & = & -\vec{U}\cdot\nabla\frac{\delta n}{n}-\nabla\cdot\delta\vec{u} \nonumber \\
&  & -\delta\vec{u}\cdot\nabla\ln n - \nabla\cdot\left(\frac{\delta n}{n}\delta\vec{u}\right)-\frac{\delta n}{n}\left(\nabla\cdot\vec{U}+\vec{U}\cdot\nabla\ln n\right) + \mathcal{O}\left(\rho_*^2\right),
\end{eqnarray}
where we have dropped all terms $\sim\mathcal{O}(\rho_*^2)$ and smaller.  The first two terms on the right-hand-side are $\sim\mathcal{O}(v_{th}/a)$ and the following three terms are $\sim\mathcal{O}(\rho_*v_{th}/a)$.  Note that we have not yet made any assumptions about the nature of the mean flow $\vec{U}$ (beyond it being large-scale) or about time scale of the fluctuations.
\newline\indent
In fact, the $\rho_*$ ordering, which is the standard gyrokinetic ordering \cite{frieman_pf_1982}, can take us further: it is possible to show that compressibility effects are order $\rho_*$, i.e., $\nabla\cdot\delta\vec{u}\sim\mathcal{O}(\rho_*)$, and that the mean flow to lowest order is purely toroidal \cite{hirshman_nf_1981, helander_cam_2002}: $\vec{U}=U_z\hat z+\vec{U}_1$, where $z$ is the toroidal direction (locally) and $\vec{U}_1\sim\mathcal{O}(\rho_*)$ including all poloidal flows\footnote{Note that the poloidal velocity $U_y$ of the bulk plasma ions has been measured with the CXRS system to be only a few $km/s$ on MAST \cite{field_ppcf_2009}, which is consistent with $U_y\sim\mathcal{O}(\rho_*)$.  Such measurements are, however, not routinely available for MAST, and one of the goals for this study is to confirm that $U_y$ is indeed small.} and first-order corrections to $U_z$ (radial flows, associated with particle fluxes, are, in fact, even smaller).  Coupled with the fact that mean quantities have no toroidal variation in a tokamak, this means that that the fifth term on the right-hand-side of equation (\ref{eq:continuity_equation_ordered}) is also $\sim\mathcal{O}(\rho_*^2)$, while the first term can be expressed as
\begin{eqnarray}\label{eq:delta_n_along_U}
\vec{U}\cdot\nabla\frac{\delta n}{n}&=&U_z\frac{\partial}{\partial z}\frac{\delta n}{n} + \vec{U}_1\cdot\nabla\frac{\delta n}{n} \nonumber \\
&=&-U_z\frac{b_y}{b_z}\frac{\partial}{\partial y}\frac{\delta n}{n}+\frac{U_z}{b_z}\hat b\cdot\nabla\frac{\delta n}{n}+\vec{U}_1\cdot\nabla\frac{\delta n}{n},
\end{eqnarray}  
where $\hat b=(0, b_y, b_z)$ is the unit vector in the direction of the magnetic field in a local orthogonal Cartesian system ($x$: radial, $y$: poloidal and $z$: toroidal), and we have used the identity $\mbox{ $\hat{b}\cdot\nabla=b_y\partial/\partial y + b_z\partial/\partial z$}$.  Making a further assumption, again standard in gyrokinetics, that the parallel spatial scale of the fluctuating quantities is $\sim\mathcal{O}(a)$, we conclude that the second term in the second line of equation (\ref{eq:delta_n_along_U}) is $\mathcal{O}(\rho_*)$.
\newline\indent
Finally, combining equation (\ref{eq:delta_n_along_U}) with (\ref{eq:continuity_equation_ordered}), we find 
\begin{equation}\label{eq:continuity_equation_BES}
\frac{\partial}{\partial t}\frac{\delta n}{n} + U_{eff}\frac{\partial}{\partial y}\frac{\delta n}{n}=\gamma\frac{\delta n}{n},
\end{equation}
where $U_{eff}=-U_zb_y/b_z=-U_z\tan\alpha$ is the dominant apparent velocity of the density patterns ($\alpha$ is the local pitch angle of the magnetic field line).  The term containing $U_{eff}$ is the only $\mathcal{O}(\rho_*^0)$ term in equation (\ref{eq:continuity_equation_BES}).  The $\mathcal{O}(\rho_*)$ and higher terms have been assembled in the right-hand-side: by definition, $\gamma$ is such that 
\begin{eqnarray}\label{eq:definition_gamma}
\gamma\frac{\delta n}{n}&=&-\frac{U_z}{b_z}\hat b\cdot\nabla\frac{\delta n}{n}-\vec{U_1}\cdot\nabla\frac{\delta n}{n} \nonumber \\
& &- \nabla\cdot\delta\vec{u}-\delta\vec{u}\cdot\nabla\ln n - \nabla\cdot\left(\frac{\delta n}{n}\delta\vec{u}\right) + \mathcal{O}(\rho_*^2).
\end{eqnarray}
This contains, in order of terms, the effects associated with
\newline\noindent
(1) parallel variations of the fluctuations,
\newline\noindent
(2) mean poloidal flows of bulk plasma ions,
\newline\noindent
(3) compressibility of the fluctuations,
\newline\noindent
(4) linear response to mean density gradient (drift waves),
\newline\noindent
(5) nonlinear effects (turbulence),
\newline\noindent
and a slew of higher-order effects of varying degree of obscurity.
\newline\indent
Thus, the right-hand-side of equation (\ref{eq:continuity_equation_BES}) contains all the nontrivial physics of waves and turbulence in the plasma.  The apparent velocity of the density patterns detected by the 2D BES system will not be influenced by these effects to dominant order --- if the orderings assumed above are correct.  What it does contain is the poloidal signature $U_{eff}$ of the dominant toroidal rotation of the plasma --- the `rotating barber-pole' effect discussed at the beginning of this section.  Indeed, if equation (\ref{eq:continuity_equation_BES}) holds and its right-hand-side is small, then, to lowest order, the density patterns just drift in the $y$-direction (poloidal) with the velocity $U_{eff}$, so the maximum of the cross-correlation function (\ref{eq:cc_definition}) will be achieved at $\tau=\Delta y/U_{eff}$.  Hence equation (\ref{eq:simple_v_bes_relation}) for the BES-measured velocity.
\newline\indent
If we are able to confirm equation (\ref{eq:simple_v_bes_relation}) experimentally, this means that the theoretical considerations employed above are consistent with the experiment.  This is important because most of the theories of tokamak turbulence rely on such considerations.  Note that there are no separate diagnostics capable of measuring individually all the $\mathcal{O}(\rho_*)$ terms in equation (\ref{eq:definition_gamma}).  Therefore, the only conclusion one can formally draw from equation (\ref{eq:simple_v_bes_relation}) holding is that the sum of these terms is small.

\section{Experimental results}\label{sec:exp_results}
\begin{figure}[!t]
\centering
\includegraphics[width=5.5in]{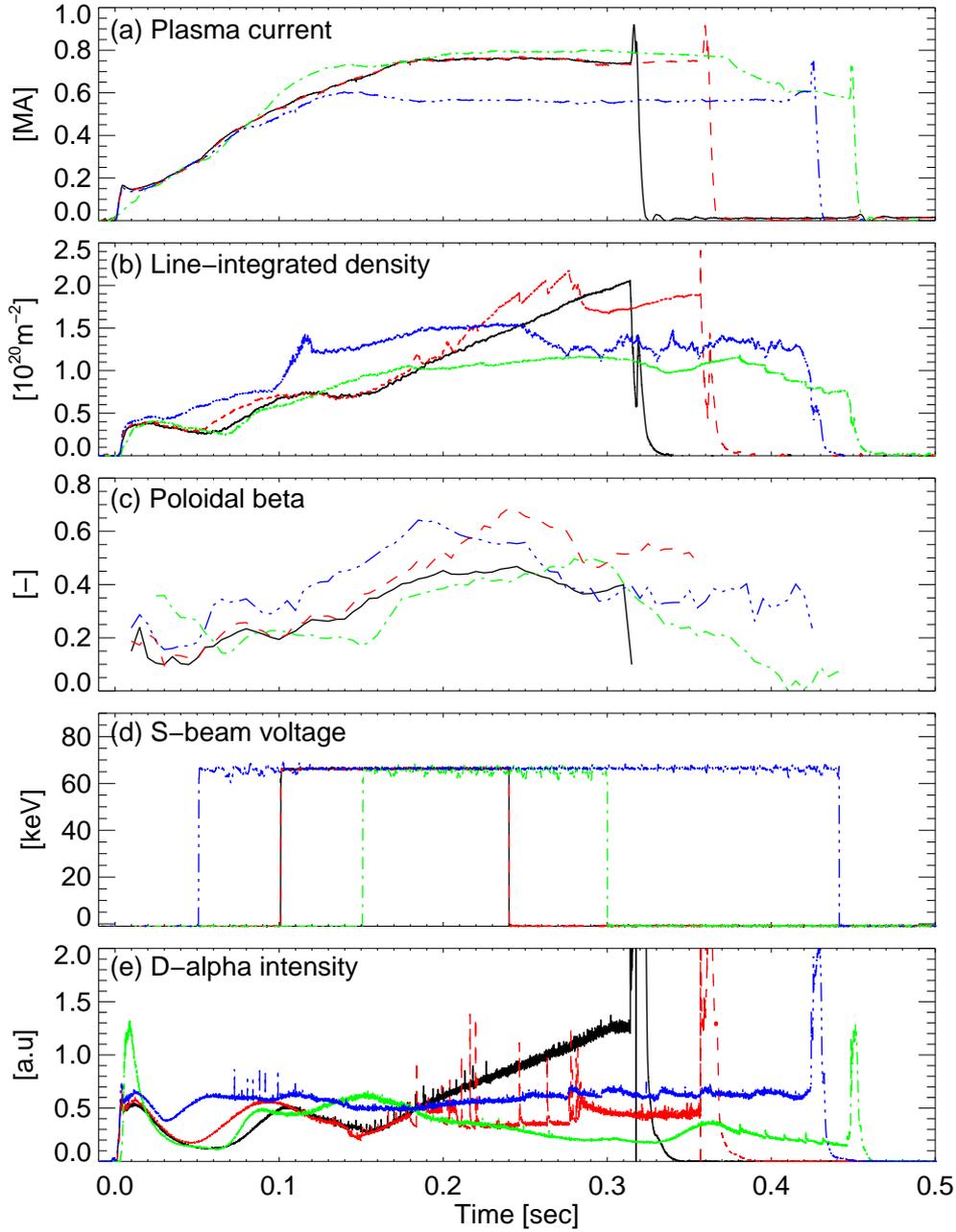}
\caption{\label{fig:basic_info}  Evolution of (a) plasma current (b) line-integrated electron density (c) poloidal beta (d) NBI (S-beam) injection energy and (e) edge $D_\alpha$ intensity of L-mode (shot \#27278, black solid),  H-mode (shot \#27276, red dash), ITB (shot \#27269, green dash dot) and high-poloidal-beta (shot \#27385, blue dash dot dot) discharges.}
\end{figure}

In this section, we apply the CCTD method to 2D BES data from MAST discharges to determine the apparent mean poloidal motion ($v_y^{BES}$) of the ion density patterns.  Then, $v_y^{BES}$ is compared with the `rotating barber-pole' velocity ($U_z\tan\alpha$) where the toroidal plasma velocity $U_z$ is obtained from the CXRS system \cite{conway_rsi_2006} and the local magnetic pitch angle $\alpha$ either from \texttt{EFIT} equilibrium reconstruction \cite{lao_nf_1985} or the MSE system \cite{kuldkepp_rsi_2006, debock_rsi_2008}.
\newline\indent
The 2D BES data are first bandpass-filtered from $20.0$ to $100.0\:kHz$ to reduce the noise level.  The low-pass filter removes the high-frequency noise component from the photon shot noise and electronic noise, while the high-pass filter reduces the contribution to the signal from low-frequency, coherent MHD (magnetohydrodynamic) modes.  The apparent mean poloidal velocity of the density patterns $v_y^{BES}$ is determined from average correlation functions calculated over $25$ time intervals of $40\:\mu sec$ duration, resulting in total $1\:msec$ averaging.  Second-order polynomial fitting is applied to interpolate the correlation function on times shorter than the sampling time of $0.5\:\mu sec$ as described in \ref{sec:desc_cctd_method}.  Finally, five consecutive values of $v_y^{BES}$ obtained in this manner are averaged, so the total averaging time is $5\:msec$ which is the effective time resolution of $v_y^{BES}$.  Using these five values of $v_y^{BES}$, the time average of various errors defined in equations (\ref{eq:norm_rand_err})-(\ref{eq:norm_rand_fit_err}) in \ref{sec:def_uncertainties} are also computed.  Statistical reliability of the CCTD method is investigated in \ref{sec:the_cctd_method} by using the synthetic 2D BES data (\ref{sec:gen_syn_data}).
\newline\indent
We present measurements of $v_y^{BES}$ from four different discharges: shot \#27278 (L-mode), shot \#27276 (H-mode), shot \#27269 (ITB) and shot \#27385 (high-poloidal-beta).  All four discharges had double-null diverted (DND) magnetic configurations and co-current NBI (neutral  beam injection).  In all of these discharges, the 2D BES system viewed at nominal major radial position of $R=1.2\:m$ corresponding to normalized minor radii $r/a=0.2-0.3$ for L- and H-modes, and $r/a=0.3$-$0.4$ for ITB and high-poloidal-beta discharges.  The evolution of key parameters for these discharges is shown in Figure \ref{fig:basic_info}.  The evolution of  plasma current, line-integrated electron density and poloidal beta characterize the overall behaviour of plasmas, while the non-zero S-beam voltage corresponds to times when the 2D BES system obtains localized density fluctuation.  The $D_\alpha$ intensity trace is used to identify when the H-mode discharge (shot \#27276) goes into its H-mode: namely, at $t=0.21 - 0.28\:sec$.  Note that the ITB discharge (shot \#27269) starts developing a strong temperature gradient at $\sim\:0.2\:sec$ and the peak ion ($C^{6+}$ from the CXRS) temperature keeps increasing until the NBI cuts off at $0.3\:sec$.  The viewing position of the 2D BES system is in the middle of the strong temperature gradient region for this discharge.

\subsection{L-mode (shot \#27278), H-mode (shot \#27276) and ITB (shot \#27269) discharges: $v_y^{BES}\approx -U_z\tan\alpha$}\label{sec:exp_barberpole}
\begin{figure}[!t]
\centering
\includegraphics[width=3.5in]{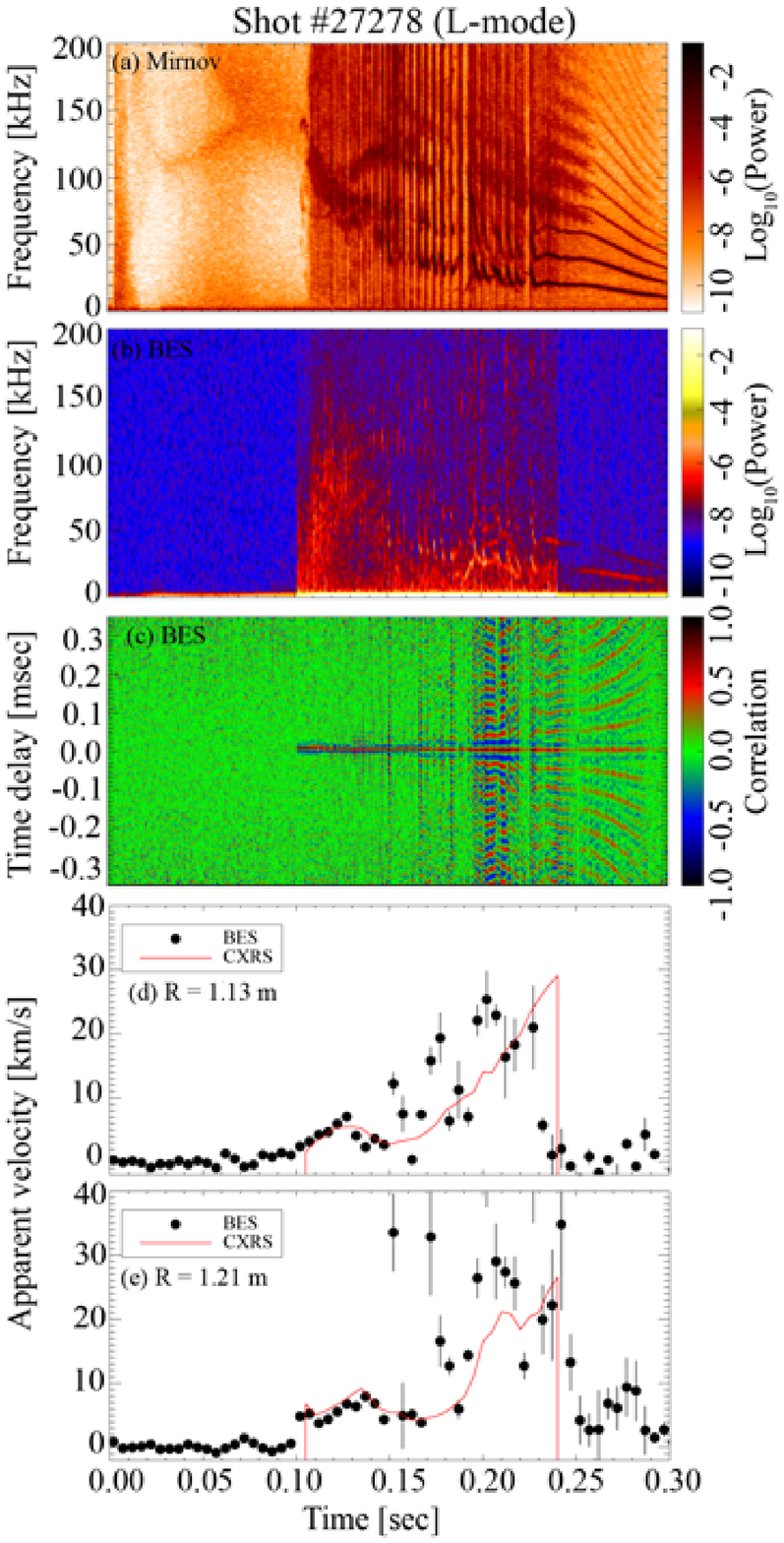}
\caption{\label{fig:27278_time_evolution} The evolution of shot \#27278 (L-mode) showing (a) cross-power spectrogram of the fluctuating magnetic field signal from two toroidally separated outboard Mirnov coils, (b) cross-power spectrogram and (c) cross-correlation of the density fluctuations between two poloidally separated channels (two mid-channels separated by $2\:cm$) from BES at $R = 1.21\:m$.  The time evolution of the (minus) apparent mean poloidal velocity ($-v_y^{BES}$, circles) from BES and the `rotating barber-pole' velocity ($U_z\tan\alpha$, red solid line) from the CXRS at (d) $R = 1.13\:m$ and (e) $R = 1.21\:m$.  BES signals are bandpass-filtered from $20.0-100.0\:kHz$ for (c)-(e).}
\end{figure}

\begin{figure}[!t]
\centering
\includegraphics[width=3.5in]{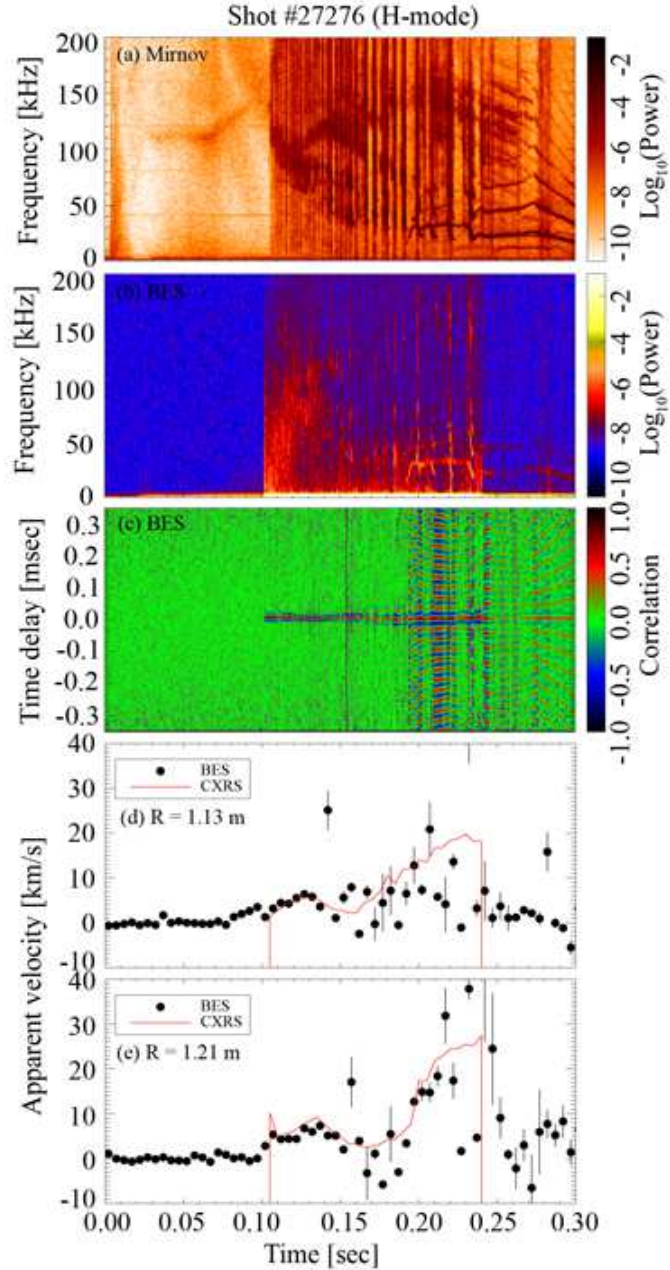}
\caption{\label{fig:27276_time_evolution} Same as Figure \ref{fig:27278_time_evolution} for shot \#27276 (H-mode).}
\end{figure}

\begin{figure}[!t]
\centering
\includegraphics[width=3.5in]{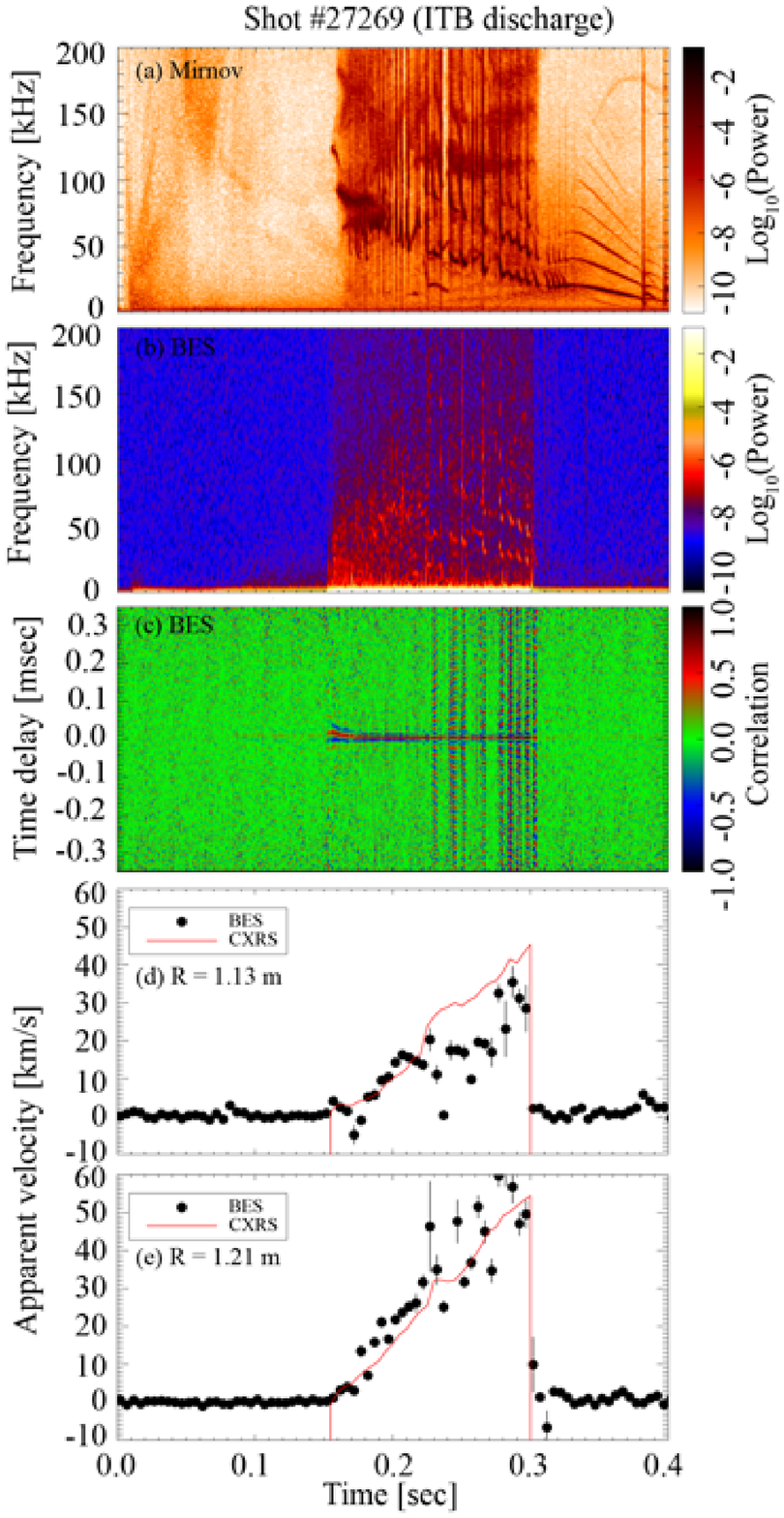}
\caption{\label{fig:27269_time_evolution} Same as Figure \ref{fig:27278_time_evolution} for shot \#27269 (ITB discharge).}
\end{figure}

Time evolution of (a) cross-power of the fluctuating magnetic field signal from two toroidally separated outboard Mirnov coils, (b) cross-power and (c) temporal cross-correlation of density fluctuations from two poloidally separated BES channels (two mid-channels separated by $2\:cm$) located at $R = 1.21\:m$ are shown in Figures \ref{fig:27278_time_evolution} (L-mode), \ref{fig:27276_time_evolution} (H-mode) and \ref{fig:27269_time_evolution} (ITB discharge).  Here, a cross-power is defined as the Fourier transform (in the time domain) of the cross-correlation function (\ref{eq:cc_definition}) with finite channel separation.
The (minus) apparent mean poloidal velocity ($-v_y^{BES}$, circles) determined by the CCTD method and the `rotating barber-pole' velocity ($U_z\tan\alpha$, red solid lines) are also shown in panels (d) at $R = 1.13\:m$ and (e) at $R = 1.21\:m$ for these three discharges.  The error bars represent the mean error $\left\langle\delta v_{fit}\right\rangle$ of the least-squares fit, as discusses in \ref{sec:def_uncertainties}.
\newline\indent
Despite the fact that these three discharges belong to three very different classes, there are common features in the apparent mean poloidal velocity:
\newline\noindent
(1) $v_y^{BES}$ is not reliable (i.e., has large error bars) when strong MHD activity is present.  The cross-power spectrograms from BES show clear signatures of MHD modes with many harmonics, which hamper filtering the BES signal over the frequency domain.  The temporal cross-correlations also show that these MHD modes have much longer correlation times ($>0.3\:msec$) than the turbulent density patterns.  The effects of MHD (global) modes on the CCTD method are investigated in \ref{sec:influenc_global_mode}, where it is found that such activity can increase not only the absolute values of the bias errors but also the linear fitting errors on $v_y^{BES}$.  Thus, comparisons between $v_y^{BES}$ and $U_z\tan\alpha$ are difficult to make during the periods where the MHD activity is strong.
\newline\noindent
(2) During the periods of weak MHD activity, i.e., $0.11$-$0.15\:sec$ for the L- and H-mode discharges, and $0.16$-$0.22\:sec$ for the ITB discharge, it is clear that the apparent mean poloidal velocity of turbulent density patterns is dominated by the `rotating barber-pole' velocity, i.e., equation (\ref{eq:simple_v_bes_relation}) holds, and the sum of all the terms of the order of $\rho_*$ or higher in equation (\ref{eq:definition_gamma}) is indeed small.
\newline\indent
Note that the H-mode discharge (shot \#27276) goes into its H-mode at $\sim0.21\:sec$ (thus, $v_y^{BES}=-U_z\tan\alpha$ is only true before the L-H transition, strictly speaking), which can be seen from the $D_\alpha$ intensity trace in Figure \ref{fig:basic_info} or from the BES cross-power spectrogram in Figure \ref{fig:27276_time_evolution}: the turbulence level drops at the start of the H-mode.  Any changes of $v_y^{BES}$ during the L-H transition cannot be discussed, because the CCTD method with the current data analysis scheme is not reliable at this time due to strong MHD activity.

\subsection{High-poloidal-beta discharge (shot \#27385): $v_y^{BES} \ne -U_z\tan\alpha$}\label{sec:exp_non_barberpole}
\begin{figure}[!t]
\centering
\includegraphics[width=3.5in]{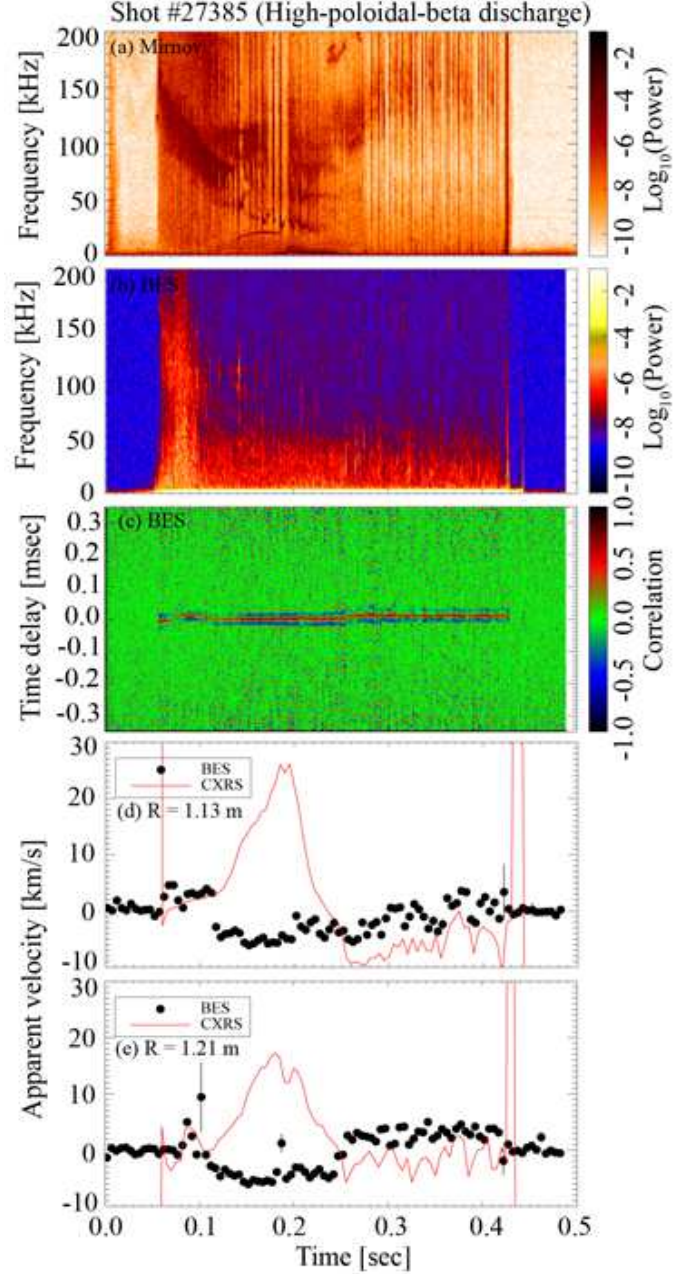}
\caption{\label{fig:27385_time_evolution} Same as Figure \ref{fig:27278_time_evolution} for shot \#27385 (high-poloidal-beta discharge).  Note that neither (b) the cross-power spectrogram nor (c) the temporal cross-correlation of the BES signal show any MHD activity.}
\end{figure}

Shot \#27385 has a relatively higher poloidal beta (the ratio of the plasma pressure to the poloidal magnetic field energy density) than the three discharges discussed in section \ref{sec:exp_barberpole} (see Figure \ref{fig:basic_info}).  Thus, it is more susceptible to tearing modes (i.e., formation of magnetic islands) \cite{haye_pop_2000, buttery_prl_2002}.  The cross-power spectrogram between the two toroidally separated outboard Mirnov coils displayed in Figure \ref{fig:27385_time_evolution}(a) shows a $m/n=3/2$ tearing mode on the $q$=1.5 flux surface starting at $\sim0.11\:sec$; its frequency increases from $<10\:kHz$ to $\sim25\:kHz$ at $\sim0.19\:sec$. 
Then, a $m/n=2/1$ mode (fundamental frequency $<10\:kHz$) develops and locks to the wall resulting in complete braking of the toroidal rotation of plasmas at $\sim0.25\:sec$.  Here, $m$ and $n$ denote the poloidal and toroidal mode numbers, respectively, and $q$ for the safety factor.
\newline\indent
 The $2/1$ mode is not expected to be seen on the BES signal as it is bandpass-filtered from $20.0-100.0\:kHz$, and no trace of the $3/2$ mode is visible in the BES signal\footnote{Because the mode flattens the mean density profile within the island, shaking of flux surfaces does not induce density fluctuations in the BES signal}.  Consequently, the $v_y^{BES}$ determined by the CCTD method does not contain large error bars during the whole discharge.
\newline\indent
The time evolution of $-v_y^{BES}$ and $U_z\tan\alpha$ in Figure \ref{fig:27385_time_evolution}(d)-(e) at two different radial locations shows that the two velocities do not agree each other at all during the period when the $3/2$ and $2/1$ modes are present.  What we find, remarkably, is that while the plasma continues to rotate toroidally (as attested by the CXRS data), there is virtually no detectable corresponding motion of the density patterns.  In fact, they seem to exhibit a weak rotation in the opposite direction to the expected rotating barber-pole effect.  Formally, this means that the plasma effects in the right-hand-side of equation (\ref{eq:continuity_equation_BES}) are not small and are able to cancel almost exactly the toroidal rotation, i.e., an effective velocity of the density patterns develops in the plasma frame that to lowest order is equal to minus the rotation velocity.  We do not currently have a theoretical explanation for this effect.  There is very little apparent difference between the turbulent density patterns in this discharge compared to others, except somewhat longer radial correlation lengths.

\section{Conclusions}\label{sec:conc}
\noindent
We have analysed 2D BES data from different types of discharges on MAST to determine the apparent mean poloidal velocities of the ion-scale density patterns using the cross-correlation time delay method.  The dominant cause of the apparent poloidal motion of the density patterns is experimentally identified to be due to the fact that field aligned patterns are advected by the background, dominantly toroidal, plasma rotational flow, i.e., the `rotating barber-pole' effect dominates the apparent mean motion of the density patterns in the lab frame.  This conclusion holds for the L-, H-mode and ITB discharges we have investigated.  An exception to this rule is found to be the investigated high-poloidal-beta discharge, where a large magnetic island is present, and the apparent velocity of the density patterns is very small, despite strong toroidal rotation.  Identifying the causes of this effect by investigating the behaviour of the turbulent density patterns quantitatively is left for future work.

\section*{Acknowledgment}
\noindent
We would like to thank Ian Abel, Steve Cowley, Edmund Highcock, Tim Horbury, Darren McDonald, Clive Michael and Jack Snape for valuable discussions, and Rob Akers for setting up the environment for CUDA programming.  This work was funded jointly by the RCUK Energy Programme, by the European Communities under the contract of Association between EURATOM and CCFE and by the Leverhulme Trust International Network for Magnetised Plasma Turbulence.  The views and opinions expressed herein do not necessarily reflect those of the European Commission.

\appendix
\renewcommand\thesection{Appendix \Alph{section}}
\section{Synthetic 2D BES data}\label{sec:gen_syn_data}
\subsection{Gaussian eddies in space and time}
\noindent
It is necessary to know the true mean velocity of the density patterns to investigate statistical reliability of the cross-correlation time delay (CCTD) method described in section \ref{sec:vel measure}.  Specifically, we investigate how reliable the CCTD method is for different magnitudes of mean velocities and correlation times of the density patterns.  We must also evaluate the effect of global modes (i.e., MHD modes) and temporally varying poloidal velocities on the CCTD method.  For this purpose, we numerically generate artificial density patterns, random both in space and time, then produce synthetic BES data using the point-spread-functions (PSFs) of the 2D BES system on MAST (as described in \ref{sec:2d_synthetic_bes_data}) and compare the inferred flow velocity with the true flow velocity.
\newline\indent
We follow a similar approach to the one suggested by Zoletnik {\it et al.} \cite{zoletnik_rsi_2005}.  Let the density patterns be described by Gaussian structures both in space and time, namely,
\begin{eqnarray} \label{eq:ghim_eddy}
\delta n\left(x, y, t \right)
	& = & \sum_{i=1}^{N} \delta n_{0i} 
		\exp \Bigg[
		-\frac{\left( x-x_{0i} \right)^2}{2\lambda_x^2} 
-\frac{\left[ y+v_y\left(t \right)\left(t-t_{0i} \right)-y_{0i}
\right]^2}{2\lambda_y^2} -\frac{\left( t-t_{0i} \right)^2}{2\tau_{life}^2} \Bigg] \times
		\nonumber \\
& & \hspace{1.45cm} \cos \Bigg[2 \pi \frac{\left[ y+v_y\left(t \right)\left(t-t_{0i}
\right)-y_{0i}
		              \right]}{\lambda_y} \Bigg],
\end{eqnarray}
where $x$, $y$ and $t$ denote radial, poloidal and time coordinates, respectively.  These numerically generated density patterns are referred to as ``eddies'' in this paper.  Here $N$ is the total number of eddies and the subscript $i$ denotes the $i^{th}$ eddy in the simulation; $\delta n_{0i}$, $x_{0i}$, $y_{0i}$ and $t_{0i}$ are the maximum amplitude and central locations in the $x$, $y$ and $t$ coordinates of the $i^{th}$ eddy, respectively; $\lambda_x$, $\lambda_y$ and $\tau_{life}$ are the widths of our Gaussian eddies in the $x$, $y$ and $t$ directions; $\tau_{life}$ is the lifetime (or the correlation time) of the eddies in the moving frame; $v_y(t)$ is the apparent advection velocity of the eddies in the poloidal direction.  Although it is possible to introduce a finite radial velocity shear by making $v_y$ a function of $x$, the effect of such shearing rates on the CCTD method is not investigated in this paper, so we will only consider $v_y$ that are independent of $x$.  The $\cos$ term in the $y$ (poloidal) direction is introduced to model wave-like-structured eddies in the poloidal direction as observed in tokamaks \cite{fonck_prl_1993}.  Note that the envelope (i.e., the $\exp$ term) and the wave structure (i.e., the $\cos$ term) of $\delta n(x, y, t)$ have the same advection velocity $v_y(t)$.  The central locations of eddies, $x_0$, $y_0$ and $t_0$, are selected from uniformly distributed random numbers, whereas their amplitudes $\delta n_0$ are selected from normally distributed random numbers whose standard deviation is one.\footnote{It is worth mentioning that there is another scheme of generating such eddies numerically, proposed by Jakubowski {\it et al.} \cite{jakubowski_rsi_2001}.  They generated the time series of fluctuating density ($\delta n_1$) using the inverse Fourier transform of a broadband Gaussian amplitude distribution in frequency space. Then, a second signal ($\delta n_2$) was generated by imposing the desired time-delay fluctuation on the $\delta n_1$ such that $\delta n_2$ was a time-delayed version of $\delta n_1$.  This method does not include spatial information for the signals.}
\newline\indent
The spatial domain of the simulation is $25\:cm$ and $20\:cm$ with the mesh size of $0.5\:cm$ in radial ($x$) and poloidal ($y$) directions, respectively. The time duration of the simulation is $20\:msec$ with a $0.5\:\mu sec$ time step so as to have the same Nyquist frequency as the real 2D BES data from MAST.  The widths $\lambda_x$ and $\lambda_y$ are set so that the full width at half maximum (FWHM) in the radial direction and the wavelength in the poloidal direction are $\sim8\:cm$ (i.e., $\lambda_x=3.53\:cm$) and $\sim20\:cm$ (i.e., $\lambda_y=20.0\:cm$), respectively, which are similar to the measured correlation lengths with the 2D BES system on MAST.\footnote{Note that Smith {\it et al.} \cite{smith_aps_2011} also reported that poloidal correlation lengths of the density patterns are $\sim20\:cm$ using their 2D BES system on NSTX.}  The eddy lifetime in the moving frame ($\tau_{life}$) is set to $15\:\mu sec$.  However, some of the data sets in this paper have different values of $\tau_{life}$, so the effect of $\tau_{life}$ on the CCTD method can be investigated.  
\newline\indent
The total number of eddies is $N=20000$.  If the eddies are too sparse in the simulation domain, then we may not achieve steady statistical results, while overly dense eddies may cause an effective widening of the specified spatial ($\lambda_x$ and $\lambda_y$) and temporal ($\tau_{life}$) correlations as many eddies can merge into one larger eddy.  Thus, we introduce another control parameter, the spatio-temporal filling factor ($F$), defined as 
\begin{equation}\label{eq:filling_factor_def}
F=N\cdot\left( \frac{\lambda_x \lambda_y}{{\rm total\;simulation\;area}}\right)\cdot\left( \frac{\tau_{ac}}{{\rm total\;simulation\;time}} \right),
\end{equation} 
where $\tau_{ac}$ is the autocorrelation time calculated as 
\begin{equation}
\tau_{ac}=\frac{\tau_{life}\left(\lambda_y/v_y\right)}{\sqrt{\tau_{life}^2+\left(\lambda_y/v_y\right)^2}}
\end{equation}
for the generated eddies defined by equation (\ref{eq:ghim_eddy}).  All of our synthetic data was generated so as to $F\sim\mathcal{O}(1)$.
\newline\indent
The testing of the CCTD method will involve exploiting what happens if $v_y(t)$ has a mean and a temporally varying components.  Thus, we generate a temporal structure of $v_y$: at each $x$,
\begin{eqnarray}\label{eq:gam_generator}
v_y\left(t \right) & = & \left\langle v_y \right\rangle + \delta v_y \left(t \right) \nonumber \\
                           & = & \left\langle v_y \right\rangle + \tilde{v}_y\left(t\right) \ast \exp \left[-\frac{t^2}{\tau_{fluc}^2} \right]\sin \left(2\pi f_{fluc}t \right)
\end{eqnarray}
where $\left\langle v_y \right\rangle$ and $\delta v_y$ are the mean and temporally varying velocities, respectively, $\tau_{fluc}$ and $f_{fluc}$ are the lifetime and frequency of $\delta v_y(t)$, respectively, and $\tilde{v}_y(t)$ is generated from normally distributed random numbers.  The RMS (root-mean-square) value of $\delta v_y(t)$ denoted as $\delta v_y^{RMS}$ will be varied as well as $\left\langle v_y \right\rangle$ to investigate the effects of these quantities on the CCTD method.  $\tau_{fluc}$ and $f_{fluc}$ allow one to introduce structured temporally varying velocities, while the randomness is kept by $\tilde{v}_y$.  As one of the causes for the temporal variation of the poloidal velocity is believed to be the existence of geodesic acoustic modes (GAMs)\footnote{We do not investigate whether the CCTD method is able to detect such a temporally structured $\delta v_y\left(t\right)$ (or GAMs) in this paper, rather we investigate how the existence of these structures affects the CCTD-determined mean velocity.} \cite{winsor_pf_1968}, we choose $\tau_{fluc}=500\:\mu sec$ and $f_{fluc}=10\:kHz$ to mimic the GAM features detected by Langmuir probes on MAST \cite{robinson_prl_2011}.
\newline\indent
The simulations have been run on a NVIDIA\textregistered\space GeForce GTS 250 GPU card using CUDA programming, which increases the computational speed owing to the highly parallelizable structure of equation (\ref{eq:ghim_eddy}).

\subsection{Synthetic 2D BES data}\label{sec:2d_synthetic_bes_data}
\begin{figure}[!t]
\centering
\includegraphics[width=6.5in]{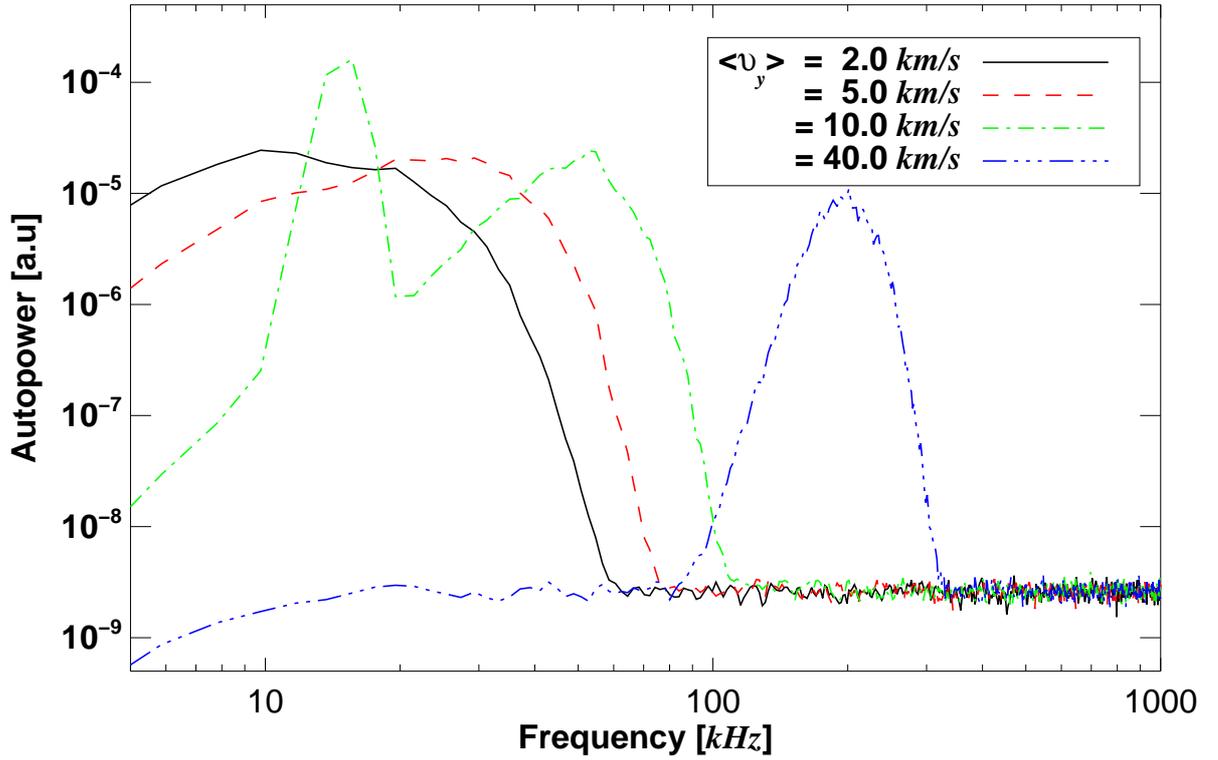}
\caption{\label{fig:syn_data_power}Autopower spectra of synthetic 2D BES data for various $\left\langle v_y \right\rangle$. Note that the spectrum for $\left\langle v_y \right\rangle = 10.0\: km/s$ (green dash dot line) has finite $I_{MHD}^{ij}$ (i.e., temporal oscillations due to global modes) in equation (\ref{eq:syn_bes_data_total_def}) at $15\:kHz$, with fluctuation level of $5\:\%$ of the DC level.  For other cases, $I_{MHD}^{RMS}=0$.  All the spectra are generated using a high-pass filter with the frequency cutoff at $5\:kHz$.}
\end{figure}

\begin{figure}[!t]
\centering
\includegraphics[width=7.0in]{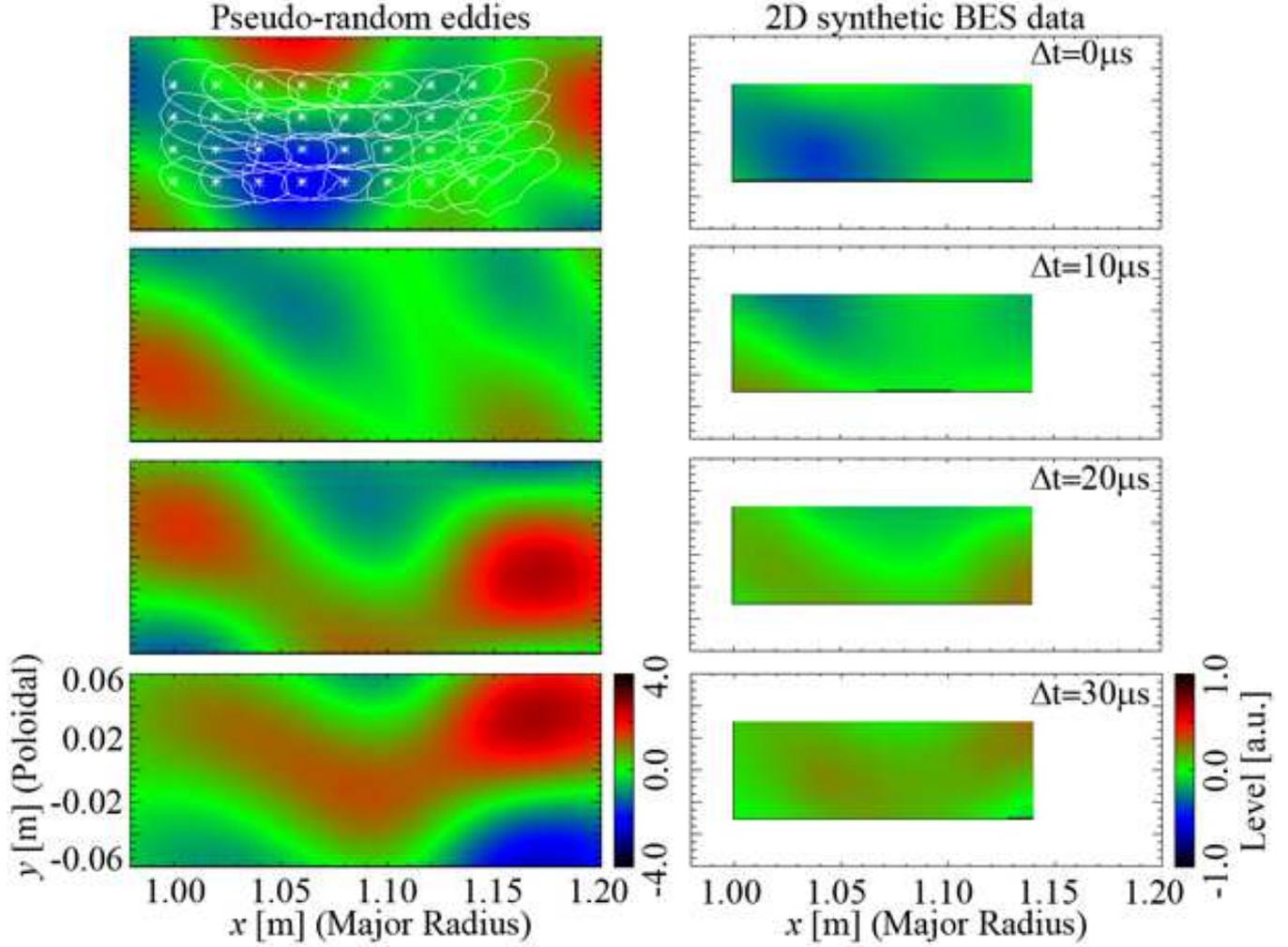}
\caption{\label{fig:syn_data_movie} Left column: four time snapshots of Gaussian eddies, $\delta n\left( x, y, t \right)$ given by equation (\ref{eq:ghim_eddy}).  Right column: the corresponding normalized synthetic 2D BES data given by equation (\ref{eq:syn_bes_data_total_def}) without the DC component.  White lines in the top left panel show the $1/e$ contour lines of the PSFs \cite{ghim_rsi_2010}, and the white asterisks show the optical focal points of the 32 channels of the 2D BES system.}
\end{figure}

\noindent
We generate the $i^{th}$ (1 to 8) radial and $j^{th}$ (1 to 4) poloidal channel of the synthetic BES data $I^{ij}\left(t \right)$ by using the calculated point-spread-functions (PSFs) of the actual 2D BES system on MAST \cite{ghim_rsi_2010} and $\delta n\left(x, y, t \right)$ from equation (\ref{eq:ghim_eddy}) with an additional random noise.  Furthermore, a large-scale (in space) coherent (in time) oscillation is included to imitate a global MHD mode.  Namely, $I^{ij}\left( t \right)$ is defined as
\begin{equation}\label{eq:syn_bes_data_total_def}
I^{ij}\left(t \right)=I_{DC}^{ij} + \delta I^{ij}\left(t \right) + I_{MHD}^{ij}\left(t\right) + I_N^{ij}\left(t \right),
\end{equation} 
where $I_{DC}^{ij}$ is the DC value -- a typical value of $0.8\:V$ is used for all channels \cite{field_rsi_2011}.  The rest of the terms are as follows.  
\newline\indent
$\delta I^{ij}\left(t\right)$ is the fluctuating part of the signal generated from the Gaussian eddies, $\delta n\left(x,y,t\right)$, given by equation (\ref{eq:ghim_eddy}) and convolved with the PSFs of the 2D BES system :
\begin{equation}\label{eq:syn_bes_data_fluc_def}
\delta I^{ij}\left(t \right)=\delta I^{RMS} \int\int \delta n\left(x,y,t \right)\mathcal{P}^{ij}\left(x, y \right)\,\mathrm{d}x\mathrm{d}y,
\end{equation}
where $\mathcal{P}^{ij}\left(x,y\right)$ is the PSF of the $i^{th}$ and $j^{th}$ channel of the 2D BES system, normalized so that RMS value of $\delta I^{ij}\left(t\right)$ is $\delta I^{RMS}$.  This value is set so that the ratio of $\delta I^{RMS}$ to $I_{DC}^{ij}$ is 0.05.  An example of the PSFs for the 32 channels of the 2D BES system on MAST is shown in Figure \ref{fig:syn_data_movie} (white contour lines in the top-left panel).
\newline\indent
$I^{ij}_{MHD}\left(t\right)$ models an MHD (global) mode.  We assume that the spatial scale of the MHD modes is larger than the BES domain in the poloidal direction, so $I^{ij}_{MHD}\left(t\right)$ does not vary in the poloidal direction.  The model MHD signal is generated in a way similar to temporal behaviour of $v_y\left(t\right)$ using equation (\ref{eq:gam_generator}), except that the mean value of $I^{ij}_{MHD}\left(t\right)$ is zero and $\tau_{fluc}=250\:\mu sec$.  The frequency of the mode $f_{MHD}$ and its RMS value, denoted $I_{MHD}^{RMS}$, will be varied in various tests.  The value of $\tau_{fluc}$ here is representative of MHD burst-like fishbone instabilities \cite{mcguire_prl_1983} or chirping modes \cite{gryaznevich_ppcf_2004} in tokamaks, for which the spectrum has a finite bandwidth. 
\newline\indent
$I_N^{ij}\left(t\right)$ represents the noise in the signal.  As the noise of the 2D BES system on MAST is dominated by the photon noise \cite{dunai_rsi_2010}, $I_N^{ij}\left(t\right)$ is generated using normally distributed random numbers.  Its RMS level is set such that the signal-to-noise ratio ($SNR$) is $300$, which is typical of the 2D BES system on MAST \cite{field_rsi_2011}.
\newline\indent
Figure \ref{fig:syn_data_power} shows examples of autopower spectra of the synthetic 2D BES data for $\left\langle v_y \right\rangle=2.0, 5.0, 10.0$ and $40.0\:km/s$.  The autopower spectrum is calculated as $|FT\{I^{ij}\left(t \right)\}|^2$ where $FT\left\{\cdot\right\}$ is the Fourier transform in the time domain.
Increasing the value of $\left\langle v_y \right\rangle$ has two effects: Doppler shift and broadening of the spectra, as expected.  Note that in Figure \ref{fig:syn_data_power}, the data for $\left\langle v_y \right\rangle=10.0\:km/s$ contains the finite $I_{MHD}^{RMS}$ with $f_{MHD}=15\:kHz$ and $I_{MHD}^{RMS}/I_{DC}^{ij}=0.05$, while $I_{MHD}^{RMS}=0$ for other cases.
\newline\indent
Figure \ref{fig:syn_data_movie} shows several time snapshots of artificial Gaussian eddies (equation (\ref{eq:ghim_eddy})) in the left column and the corresponding synthetic 2D BES data in the right column (with DC component removed from equation (\ref{eq:syn_bes_data_total_def})).
The eddies are moving upward with $\left\langle v_y \right\rangle = 5.0\:km/s$.  The top left panel in this figure also shows the $1/e$ contour lines of the PSFs for the 32 channels \cite{ghim_rsi_2010}. Snapshots for the synthetic 2D BES data are generated with the bandpass frequency filtering from $10$ to $70\:kHz$ to suppress the noise. As the synthetic 2D BES data have only 32 spatial points, spatial interpolation is performed using parametric cubic convolution technique \cite{park_image_processing_1982}.

\section{Assessment of the CCTD method}\label{sec:the_cctd_method}
\noindent
In this section, errors involved in determining the mean velocity of the density patterns by the CCTD method are examined using the synthetic 2D BES data generated according to the procedure explained in \ref{sec:gen_syn_data}. The velocity measured via the correlation function (equation (\ref{eq:cc_definition})) is denoted $v_y^{BES}$ and compared with the prescribed value $\left\langle v_y \right\rangle$ that appears in equation (\ref{eq:gam_generator}), i.e., the mean poloidal velocity of the synthetic data.  \ref{sec:desc_cctd_method} provides detailed description of the CCTD method used in this paper, then four types of error are identified for the quantitative comparisons.  These errors are evaluated in \ref{sec:mean_vel_detect} and \ref{sec:influence_tau_life} for different values of $\left\langle v_y \right\rangle$ and the eddy correlation time $\tau_{life}$.  Subsequent sections are devoted to investigating how the existence of global (MHD) modes and temporally varying poloidal velocity affect the errors.

\subsection{Description of the CCTD method}\label{sec:desc_cctd_method}
\noindent
As defined by equation (\ref{eq:cc_definition}), cross-correlation functions are calculated as time averages of the data.  For a $20 \:msec$-long synthetic data set containing $N_{total}=40,000$ data points with the sampling time $\Delta t_{sam}=0.5\:\mu sec$, we want to determine $v^{BES}_y$ with a time resolution $t_{res}=1\:msec$.  First, a cross-correlation function (\ref{eq:cc_definition}) is calculated on a sub-time window of the synthetic 2D BES data containing $N_{\mathcal{C}}$ points, where $N_{\mathcal{C}} < t_{res}/\Delta t_{sam}$.  Then, such cross-correlation functions are averaged over $N_{avg}$ consecutive sub-time windows where $N_{avg}=(t_{res}/\Delta t_{sam})/N_{\mathcal{C}}$ so that an averaged cross-correlation function is obtained at every $t_{res}$.  In this paper, we use $N_{\mathcal{C}}=80$, so $N_{avg}=25$.
\newline\indent
Denoting $f(t)$ and $g(t)$ the time series over a sub-time window from two poloidally separated synthetic 2D BES channels, the cross-correlation function (\ref{eq:cc_definition}) for this sub-time window is:
 \begin{equation}\label{eq:modified_cc_def}
\mathcal{C}_{sub}\left(r\Delta t_{sam}\right)=\frac{\frac{1}{N_{\mathcal{C}}}\displaystyle\sum_{k=0}^{N_{\mathcal{C}}-1}f\left(k\Delta t_{sam}\right)\:g\left((k+r)\Delta t_{sam}\right)}{\frac{1}{N_{\mathcal{C}}-1}\sqrt{\displaystyle\sum_{k=0}^{N_{\mathcal{C}}-1}f^2\left(k\Delta t_{sam}\right)\displaystyle\sum_{k=0}^{N_{\mathcal{C}}-1}g^2\left(\left(k+r\right)\Delta t_{sam}\right)  }},
\end{equation}
for any integer $r$ with $|r| < N_{\mathcal{C}}-1$.  Finally, by averaging $\mathcal{C}_{sub}$ for $N_{avg}$ consecutive sub-time windows we obtain the smoothed averaged cross-correlation function $\mathcal{C}(r\Delta t_{sam})$ from $1\:msec$-long data points.
\newline\indent
The CCTD method has a serious limitation due to the fact that the sampling time $\Delta t_{sam}$ is finite.  In order to calculate $v_y^{BES}$  using only two poloidally separated channels, a line is fitted through two points on a $\left(\Delta y,\tau_{peak}^{cc}\right)$ plane as shown in Figure \ref{fig:cctd_method}(b).  The first point is located at $\left(\Delta y, \tau_{peak}^{cc}\right)=\left(0,0\right)$ by definition, and the second point at $\left(\Delta y, r\:\Delta t_{sam}\right)$.  Then, possible values of $v_y^{BES}$ are restricted to $\Delta y/\left(r\Delta t_{sam}\right)$ where $r$ is an integer.  For the 2D BES system on MAST, using two adjacent poloidal channels ($\Delta y=2.0\:cm$) with a sampling time $\Delta t_{sam}=0.5\:\mu sec$, the possible values of $v_y^{BES}$ are limited to $40.0,\:20.0,\:13.3,\ldots\:km/s$ for $r=1,\:2,\:3,\ldots$.  Such a limitation may be mitigated by using four poloidally separated channels.  However, using four channels is not always possible if the channels that are farthest apart are not correlated.  To resolve this issue, we use a second-order polynomial fit on the cross-correlation function $\mathcal{C}(r\Delta t_{sam})$ to locate its global maximum: if $r_{peak}$ is the point where the discrete cross-correlation function $\mathcal{C}(r\Delta t_{sam})$ is maximum, we use the three values of $\mathcal{C}(r\Delta t_{sam})$ at $r=r_{peak}$, $r_{peak}-1$ and $r_{peak}+1$ to fit a second-order polynomial.  The ``true'' maximum is found from this fit.  We denote the time delay at which this maximum is reached by $\tau_{peak}^{cc}$.

\subsection{Definition of errors}\label{sec:def_uncertainties}
\noindent
For a given set of $20\:msec$-long synthetic 2D BES data, we calculate $v_y^{BES}$ with the time resolution of $1\:msec$ (\ref{sec:desc_cctd_method}).  Furthermore, we do this at three different radial locations\footnote{As described in \ref{sec:gen_syn_data}, $v_y\left(t\right)$ are identical at all radial locations.  One column in the middle and two columns from the edges of the 2D BES channels are used.} so that the average of $v_y^{BES}$, denoted $\left\langle v_y^{BES}\right\rangle$, can be calculated using $60$ values of $v_y^{BES}$.  To make quantitative comparisons between $\left\langle v_y^{BES}\right\rangle$ and $\left\langle v_y \right\rangle$ defined in equation (\ref{eq:gam_generator}), we define four types of error.
\newline\indent
The normalized bias error
\begin{equation}\label{eq:norm_bias_err}
\hat\sigma_{bias}=\frac{\left\langle v_y^{BES} \right\rangle - \left\langle v_y\right\rangle}{\left\langle v_y \right\rangle}
\end{equation}
is a quantitative measurement of the systematic discrepancy between the measured and the true value.  The normalized random error
\begin{equation}\label{eq:norm_rand_err}
\hat\sigma_{rand}=\frac{\sqrt{\left\langle\left( v_y^{BES}-\left\langle v_y^{BES}\right\rangle \right)^2 \right\rangle}}{\left|\left\langle v_y^{BES} \right\rangle\right|}
\end{equation}
quantifies the degree of fluctuation in the measured $v_y^{BES}$ with respect to $\left\langle v_y^{BES} \right\rangle$. This value may depend on the MHD contribution in equation (\ref{eq:syn_bes_data_total_def}) and the temporally varying poloidal velocity $\delta v_y(t)$ in equation (\ref{eq:gam_generator}).
\newline\indent
Furthermore, as linear fitting is done to determine $v_y^{BES}$ (see Figure \ref{fig:cctd_method}), two other types of error are present.  The slope of a linear fit can be denoted as $v_y^{BES} \pm \delta v_{fit}$ where $\delta v_{fit}$ is a degree of the uncertainty of the least-square fit\footnote{In Figure \ref{fig:cctd_method}, we plotted $\tau_{peak}^{cc}$ as a function of $\Delta y$ and determined $v_y^{BES}$ as the inverse of the slope of a fitted line.  Operationally, we actually plot $\Delta y$ as a function of $\tau_{peak}^{cc}$ so the slope of a fitted line is the $v_y^{BES}$.}.  Then, the normalized mean of $\delta v_{fit}$ is
\begin{equation}\label{eq:norm_mean_fit_err}
\hat\sigma_{mean}^{fit}=\frac{\left\langle \delta v_{fit} \right\rangle}{\left|\left\langle v_y^{BES} \right\rangle\right|},
\end{equation}
and the normalized random error in $\delta v_{fit}$ is
\begin{equation}\label{eq:norm_rand_fit_err}
\hat\sigma_{rand}^{fit}=\frac{\sqrt{\left\langle\left(\delta v_{fit}-\left\langle\delta v_{fit}\right\rangle\right)^2\right\rangle}}{\left|\left\langle v_y^{BES} \right\rangle\right|}.
\end{equation}
These two uncertainties together provide an estimation of how well a linear line is fitted to given data points.  For example, if the assumption that $\tau_{life}$ is long enough so that all four poloidally separated channels observe the same eddies is not satisfied, then $\hat\sigma_{mean}^{fit}$ becomes large.  On the other hand, if this assumption is occasionally satisfied, then $\hat\sigma_{rand}^{fit}$ exhibits such events because $\delta v_{fit}$ will then be small compared to its average.  Note that error bars of the CCTD-determined apparent velocities in Figures \ref{fig:27278_time_evolution} - \ref{fig:27385_time_evolution} show $\left\langle \delta v_{fit} \right\rangle$.
\newline\indent
In the following sections, these four types of error will be evaluated for various values of $\left\langle v_y \right\rangle$ and $\tau_{life}$, and various ranges of $I_{MHD}^{RMS}$, $f_{MHD}$ and $\delta v_y^{RMS}$.

\subsection{Measuring mean velocity}\label{sec:mean_vel_detect}
\begin{figure}[!t]
\centering
\includegraphics[width=6.5in]{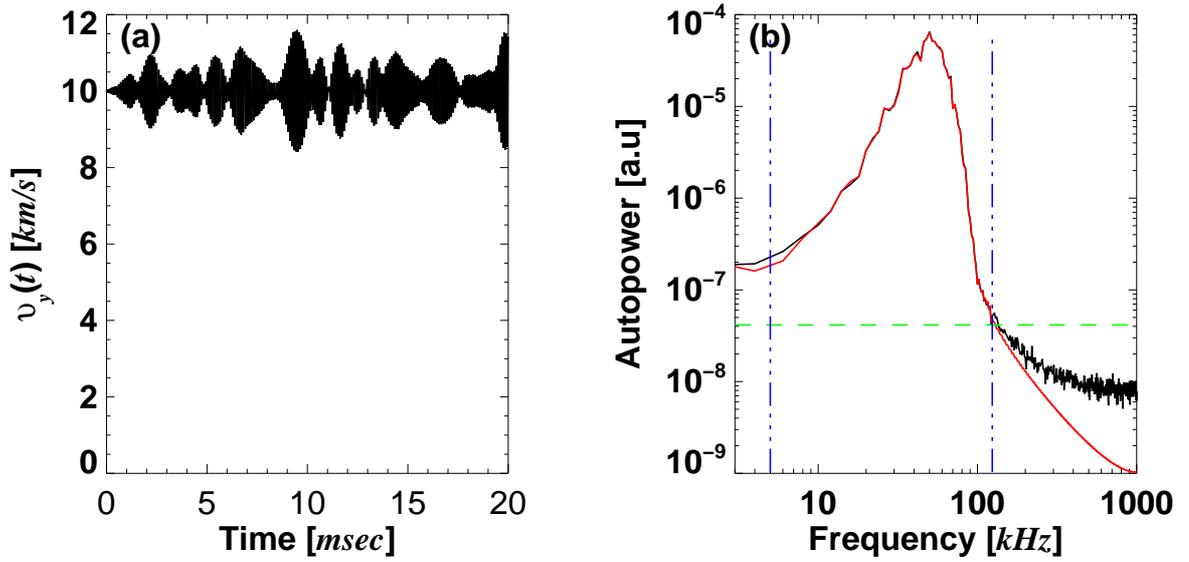}
\caption{\label{fig:vz_filtered_data} (a) Poloidal velocity $v_y\left(t \right)$ generated using equation (\ref{eq:gam_generator}) with $\left\langle v_y \right\rangle = 10.0\:km/s$.  (b) Autopower spectra of the original (black) and frequency-filtered (red) synthetic BES signals.  The green horizontal dashed line shows the noise cut-off level, defined to be $5$ times the averaged autopower level above $900\:kHz$, and vertical blue dash-dotted lines indicate the low- and high-frequency cutoffs.}
\end{figure}

\noindent
To investigate the reliability of the CCTD method described in \ref{sec:desc_cctd_method} for estimating $v_y^{BES}$, we generate a number of synthetic 2D BES data sets with various values of $\left\langle v_y \right\rangle$ while keeping all the other parameters in equations (\ref{eq:ghim_eddy}), (\ref{eq:gam_generator}), (\ref{eq:syn_bes_data_total_def}) and (\ref{eq:syn_bes_data_fluc_def}) constant.  In real experiments, there is almost always some temporal variation of $v_y$, thus the RMS value of $\delta v_y$ in equation (\ref{eq:gam_generator}) is set to $5\:\%$ of $\left\langle v_y \right\rangle$ in this subsection.  The synthetic 2D BES data are frequency-filtered to suppress the noise before the cross-correlation functions are calculated. Figure \ref{fig:vz_filtered_data} shows examples of (a) $v_y\left(t \right)$ generated according to equation (\ref{eq:gam_generator}) with $\left\langle v_y \right\rangle = 10.0\:km/s$ and (b) the original (black) and frequency-filtered (red) autopower spectra of a generated synthetic signal.  Here, the noise cut-off level is set to be the $5$ times the averaged autopower level above $900\:kHz$ (green dashed line).
\newline\indent
Figure \ref{fig:mean_vel_check} shows $\hat\sigma_{bias}$, $\hat\sigma_{rand}$, $\hat\sigma_{mean}^{fit}$ and $\hat\sigma_{rand}^{fit}$ defined in \ref{sec:def_uncertainties} and calculated for values of $\left\langle v_y \right\rangle$ ranging from $1$ to $100\:km/s$.  The basic conclusions that can be made based on these results are as follows:
\begin{figure}[!t]
\centering
\includegraphics[width=6.5in]{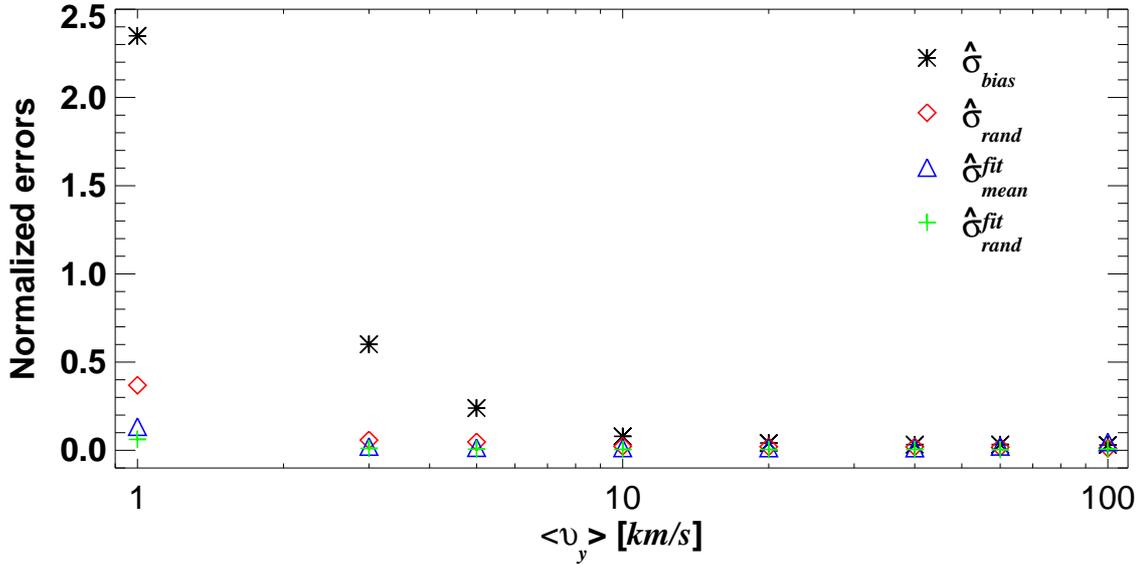}
\caption{\label{fig:mean_vel_check} The four types of error defined in equations (\ref{eq:norm_bias_err})-(\ref{eq:norm_rand_fit_err}) calculated for values of $\left\langle v_y \right\rangle$ ranging from $1$ to $100\:km/s$.}
\end{figure}
\newline\noindent
(1) For $\left\langle v_y \right\rangle \lesssim 5.0\:km/s$, the CCTD method is not reliable.  This is due to the fact that eddies do not live long enough to be detected by all the poloidally separated channels.  Indeed, it was a priori clear that $\left\langle v_y \right\rangle<\Delta y / \tau_{life}$ could not be measured. This translates to $\left\langle v_y \right\rangle<4.0\:km/s$ for $\Delta y=6.0\:cm$ and $\tau_{life}=15.0\:\mu sec$, so our results are consistent with this simple criterion.
\newline\noindent
(2) The CCTD method usually overestimates $\left\langle v_y \right\rangle$ (i.e., $\hat\sigma_{bias}>0$). This can be explained by the effective channel separation distance ($\Delta y$) being in fact slightly less than $2.0\:cm$ because of the overlapping of the PSFs, as shown in Figure \ref{fig:syn_data_movie}. 
\newline\noindent
(3) The limitation of the CCTD method due to the finite $\Delta t_{sam}$ is successfully overcome by fitting a second order polynomial to the cross-correlation function, as explained in \ref{sec:desc_cctd_method}.

\subsection{Effect of the eddy lifetime}\label{sec:influence_tau_life}
\begin{figure}[!t]
\centering
\includegraphics[width=6.5in]{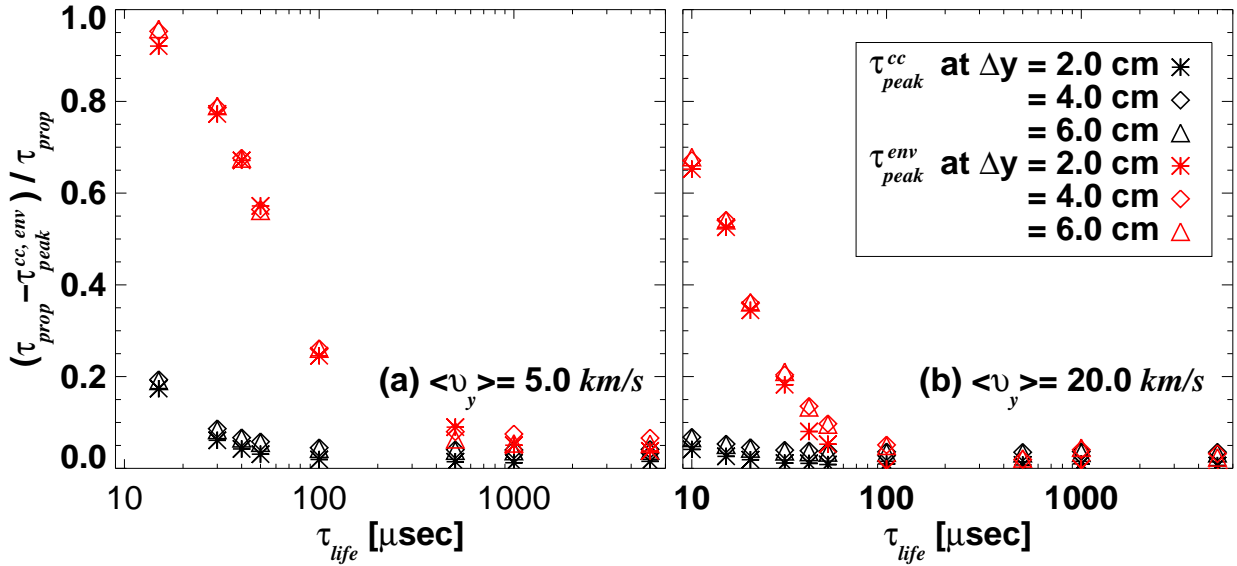}
\caption{\label{fig:lifetime_effects} Relative discrepancy between the propagation time $\tau_{prop}=\Delta y/\left\langle v_y\right\rangle$ and the times $\tau_{peak}^{cc}$ (black) or $\tau_{peak}^{env}$ (red) at which the cross-correlation function or its envelope reaches their peaks for (a) $\left\langle v_y \right\rangle=5.0\:km/s$ and (b) $20.0\:km/s$.}
\end{figure}

\noindent
As explained in section \ref{sec:physcial_meaning_cctd}, the CCTD method for determining $\left\langle v_y\right\rangle$ is based on the idea that the peak of the cross-correlation function occurs at $\tau_{peak}^{cc}=\tau_{prop}$, where $\tau_{prop}=\Delta y/\left\langle v_y\right\rangle$ is the propagation time of the fluctuating density patterns between detectors poloidally separated by the distance $\Delta y$. However, $\tau_{peak}^{cc}$ will not coincide with $\tau_{prop}$ if the lifetime $\tau_{life}$ of the fluctuations is not long compared to $\tau_{prop}$.  The failure of the method for $\left\langle v_y \right\rangle < 5.0\:km/s$ illustrated in Figure \ref{fig:mean_vel_check} is an example of what happens when $\tau_{prop}$ is too large.  Here, we investigate the effect of $\tau_{life}$ on $\tau_{peak}^{cc}$ quantitatively, via a systematic $\tau_{life}$ scan of the synthetic BES data.
\newline\indent
Two values $\left\langle v_y \right\rangle=5.0$ and $20.0\:km/s$ are chosen for this study.  For $\left\langle v_y \right\rangle=5.0\:km/s$, $\tau_{prop}=4.0$, $8.0$ and $12.0\:\mu sec$ with $\Delta y=2.0$, $4.0$ and $6.0\:cm$, respectively; for $\left\langle v_y \right\rangle=20.0\:km/s$, they are $1.0$, $2.0$ and $3.0\:\mu sec$.  The peak time $\tau_{peak}^{cc}$ is found using the polynomial fitting method described in \ref{sec:desc_cctd_method}, and $\left(\tau_{prop}-\tau_{peak}^{cc}\right)/\tau_{prop}$ as a function of $\tau_{life}$ is plotted for three different values of $\Delta y$ in Figure \ref{fig:lifetime_effects}.  It shows that $\tau_{peak}^{cc}$ underestimates the true $\tau_{prop}$ for small values of $\tau_{life}$, leading to an overestimation of the $\left\langle v_y \right\rangle$, consistent with the results shown in Figure \ref{fig:mean_vel_check}.  It is encouraging, however, that even relatively low velocities of just a few $km/s$ can be determined by the CCTD method with reasonable accuracy ($\sim\:20\%$).
\newline\indent
It is also possible to consider the global maximum of the envelope of the cross-correlation function.  We use Hilbert transform to determine the time delay $\tau_{peak}^{env}$ at which the envelope of the cross-correlation function is maximum \cite{durst_rsi_1992}.  The comparison between $\tau_{peak}^{env}$ and $\tau_{prop}$ is shown in Figure \ref{fig:lifetime_effects}.  It is clear that $\tau_{peak}^{env}$ has a much stronger dependence on $\tau_{life}$ than $\tau_{peak}^{cc}$, so this measure will not be used to estimate $\left\langle v_y \right\rangle$ in this paper. We note, however, that the strong dependence of $\tau_{peak}^{env}$ on the eddies' lifetime $\tau_{life}$ and of $\tau_{peak}^{cc}$ on their propagation time $\tau_{prop}$ may provide a way to measure correlation times in the plasma frame.  Such an investigation is currently being pursued and will be reported elsewhere.

\subsection{Effect of coherent MHD modes}\label{sec:influenc_global_mode}
\begin{figure}[!t]
\centering
\includegraphics[width=6.5in]{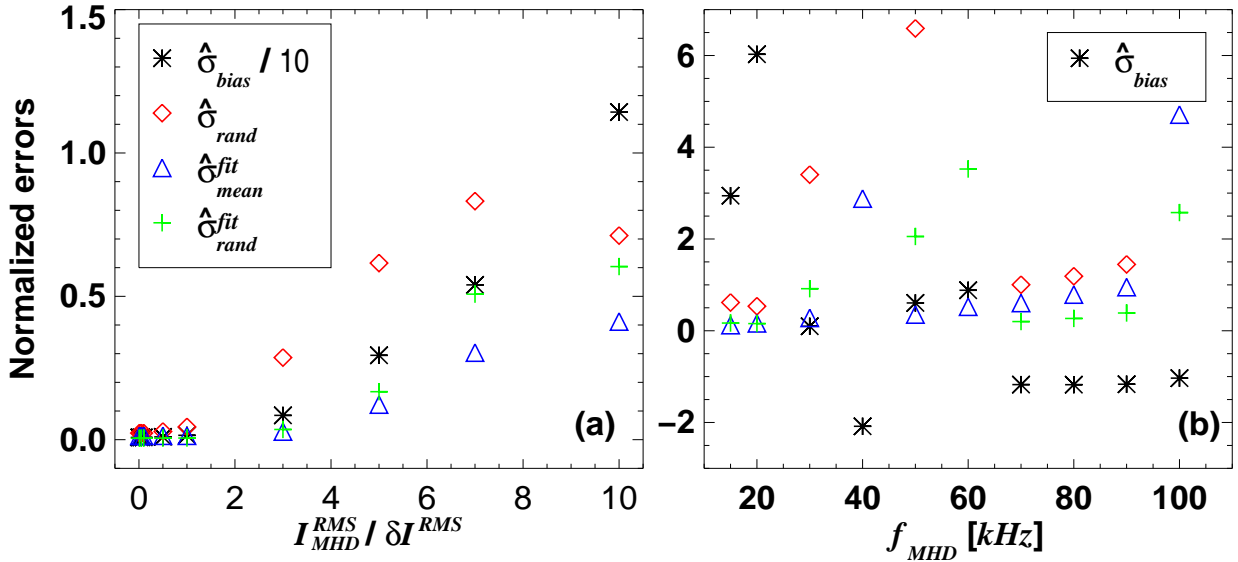}
\caption{\label{fig:check_mean_vel_with_MHD}  Four types of error (a) as functions of the RMS levels of a global mode $I_{MHD}^{RMS}$ relative to that of turbulence signal $\delta I^{RMS}$; the frequency is fixed at $f_{MHD}=15.0\:kHz$; (b) as functions of the global mode frequency $f_{MHD}$ at fixed $I_{MHD}^{RMS}/\delta I^{RMS}=5.0$.  Note that $\hat\sigma_{bias}$ in (a) is scaled down by a factor of 10, and some points are missing in (b) because they are out of the plot range.}
\end{figure}

\begin{figure}[!t]
\centering
\includegraphics[width=6.5in]{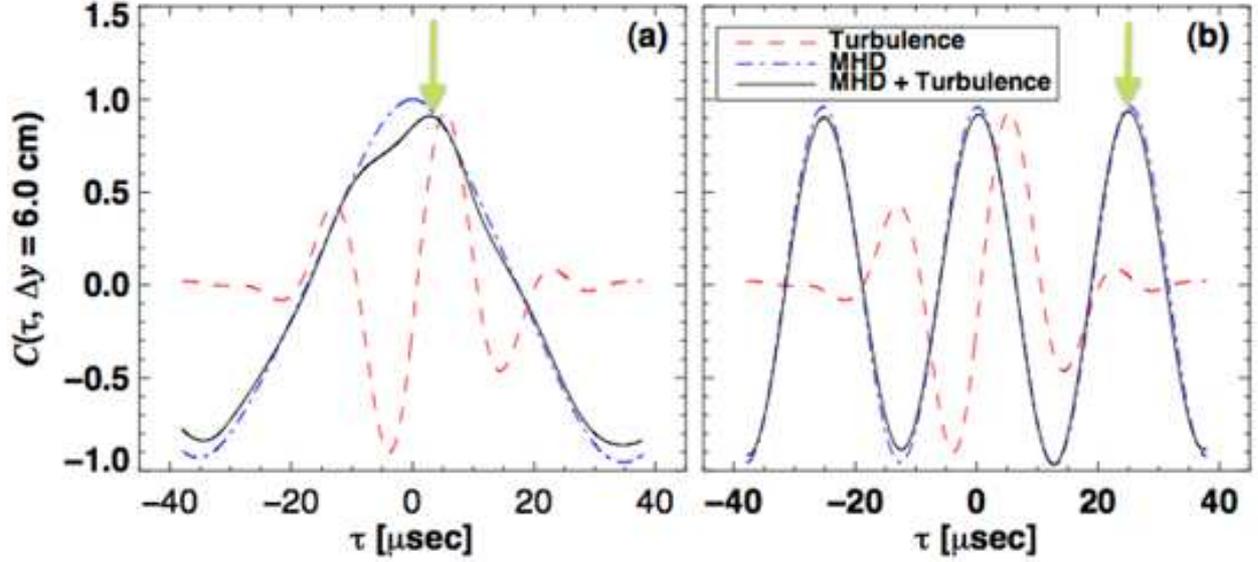}
\caption{\label{fig:cc_with_mhd} Cross-correlation functions of the random eddies only (red dash), the global mode only (blue dash dot) and the eddies with the global mode (black solid) with (a) $f_{MHD}=15.0$ and (b) $f_{MHD}=40.0\:kHz$.  Green arrows indicate the position of $\tau_{peak}^{cc}$, which does not coincide with the maximum of the cross-correlation function of the eddies only (red dash).}
\end{figure}

\noindent
Many experimental 2D BES data sets on MAST exhibit strong MHD (global mode) activity in addition to the small-scale turbulence.  Removing such global modes in the frequency domain is not straightforward as they can have multiple harmonics extending into higher frequencies.  While they could be filtered out relatively easily in the wavenumber domain, constructing wavenumber spectra with a very limited number of spatial data points is difficult.  Thus, it is useful to investigate how the presence of such modes affects the quality of our measurement of $\left\langle v_y \right\rangle$.  In this section, this is done by using synthetic BES data sets with different RMS levels $I_{MHD}^{RMS}$ and frequencies $f_{MHD}$ of the global oscillations (the $I_{MHD}^{ij}$ term in equation (\ref{eq:syn_bes_data_total_def})).
\newline\indent
The four errors ($\hat\sigma_{bias}$, $\hat\sigma_{rand}$, $\hat\sigma_{mean}^{fit}$ and $\hat\sigma_{rand}^{fit}$) are calculated for various ratio of $I_{MHD}^{RMS}$ to the RMS value of $\delta I^{ij}\left(t \right)$ (i.e., $\delta I^{RMS}$ in equation (\ref{eq:syn_bes_data_fluc_def})).  These errors are plotted in Figure \ref{fig:check_mean_vel_with_MHD}(a) for the $I_{MHD}^{RMS}$ scan.  
Here, the frequency of the global mode $f_{MHD}=15.0\:kHz$ and $\left\langle v_y \right\rangle = 10.0\:km/s$.  It is clear that if the power level of the mode is larger than that of the turbulence signal, then the CCTD method produces large bias errors $\hat\sigma_{bias}$.  To examine how the frequency of a global mode affects the errors, $f_{MHD}$ is varied with a fixed value of $I_{MHD}^{RMS}/\delta I^{RMS}=5.0$.  The results of this scan are shown in Figure \ref{fig:check_mean_vel_with_MHD}(b).  It shows that $\hat\sigma_{bias}$ can be either positive or negative with different values of $f_{MHD}$ meaning that global modes in real experimental data can cause both over- and under-estimation of the true $\left\langle v_y \right\rangle$.
\newline\indent
Figure \ref{fig:cc_with_mhd} shows how different frequencies $f_{MHD}$ can cause such an over- or under-estimation of the $\left\langle v_y \right\rangle$.  
Two identical sets of synthetic BES data with $\left\langle v_y \right\rangle=10.0\:km/s$ are generated, one with and another without a global mode at (a) $f_{MHD}=15.0\:kHz$ and (b) $f_{MHD}=40.0\:kHz$, with $I_{MHD}^{RMS}/\delta I^{RMS}=5.0$.  Without the global modes, the cross-correlation functions with $\Delta y=6.0\:cm$ (red dashes in Figure \ref{fig:cc_with_mhd}) have the expected value $\tau_{peak}^{cc}\approx 6.0\:\mu sec$ for both cases.  In contrast, the presence of the global mode in the synthetic BES data shifts $\tau_{peak}^{cc}$ towards (a) smaller time-lag (over-estimation) or (b) larger time-lag (under-estimation).  
\newline\indent
We conclude that a global (MHD) mode with $I_{MHD}^{RMS}>\delta I^{RMS}$ affects the structure of the cross-correlation functions (both the shape and the position of $\tau_{peak}^{cc}$) rendering the CCTD method unreliable.

\subsection{Effect of temporally varying poloidal velocity}\label{sec:influence_fluc_vel}
\noindent
No physical quantities are absolutely quiet in real experiments, thus it is necessary to investigate how the RMS level $\delta v_y^{RMS}$ of the temporal variation of the poloidal velocity (see equation (\ref{eq:gam_generator})) influences the measurement of $\left\langle v_y \right\rangle$. 
\begin{figure}[!t]
\centering
\includegraphics[width=6.5in]{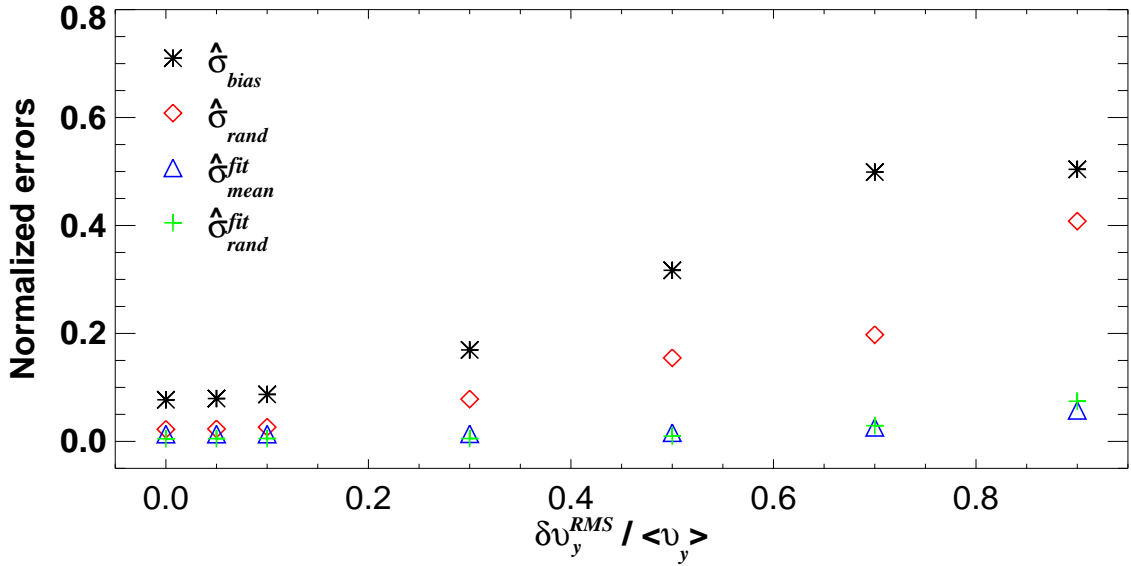}
\caption{\label{fig:check_mean_vel_with_vfluc} Four types of error for various RMS levels $\left\langle v_y\right\rangle$ of temporally varying poloidal velocities.}
\end{figure} 
\newline\indent
Figure \ref{fig:check_mean_vel_with_vfluc} shows how finite $\delta v_y^{RMS}/\left\langle v_y \right\rangle$ (with $\left\langle v_y\right\rangle=10.0\:km/s$) affect the four errors defined in \ref{sec:def_uncertainties}.  It appears that $\hat\sigma_{bias}$ saturates at around $50\:\%$ for the scenarios we have investigated, while other three errors increase without showing any sign of saturation.  Thus, the CCTD method to measure $\left\langle v_y\right\rangle$ is subject to a non-negligible bias error (up to $\sim\:50\%$) if the RMS level of temporal variation of the poloidal velocity is greater than a half of the mean poloidal velocity.

\bibliographystyle{unsrt}

\end{document}